\documentclass[11pt]{article}
\usepackage[utf8]{inputenc}
\usepackage{amsmath}
\usepackage{amsfonts}
\usepackage{amssymb}
\usepackage[english]{babel}
\usepackage{graphicx}
\usepackage[left=2cm,right=2cm,top=2.5cm,bottom=2.5cm]{geometry}
\usepackage{subfloat}
\usepackage{subcaption} 
\usepackage{float} 
\usepackage{wrapfig}
\usepackage{fancyhdr}
\usepackage{indentfirst}
\usepackage{cite}
\usepackage{physics}
\usepackage{setspace}
\usepackage{authblk}
\usepackage{xcolor}
\begin{document}

\begin{center}
\begin{LARGE}

\textbf{Microcavity phonon polaritons – from weak to ultra-strong phonon-photon coupling}
\end{LARGE}

\bigskip

\begin{large}
María Barra-Burillo$^{1,*}$, Unai Muniain$^{2,*}$, Sara Catalano$^{1}$, Marta Autore$^{1}$, Felix Casanova$^{1,3}$, Luis E. Hueso$^{1,3}$, Javier Aizpurua$^{2,4}$, Ruben Esteban$^{2,4}$ and Rainer Hillenbrand$^{3,5}$ 
\end{large}
\medskip

\begin{small}
$^1$\textit{CIC nanoGUNE BRTA, 20018 Donostia-San Sebastián, Spain}

$^2$\textit{Donostia International Physics Center, 20018 Donostia-San Sebastián, Spain}

$^3$\textit{IKERBASQUE, Basque Foundation for Science, 45011 Bilbao, Spain}

$^4$\textit{Materials Physics Center, CSIC-UPV/EHU, 20018 Donostia-San Sebastián, Spain}

$^5$\textit{CIC nanoGUNE BRTA and EHU/UPV, 20018 Donostia-San Sebastián, Spain }

*These authors contributed equally to this work.

Corresponding author: r.hillenbrand@nanogune.eu
\end{small}
\end{center}

\noindent \textbf{Abstract:} Strong coupling between molecular vibrations and microcavity modes has been demonstrated to modify physical and chemical properties of the molecular material. Here, we study the much less explored coupling between lattice vibrations (phonons) and microcavity modes. Embedding thin layers of hexagonal boron nitride (hBN) into classical microcavities, we demonstrate the evolution from weak to ultrastrong phonon-photon coupling when the hBN thickness is increased from a few nanometers to a fully filled cavity. Remarkably, strong coupling is achieved for hBN layers as thin as 10 nm. Further, the ultrastrong coupling in fully filled cavities yields a cavity polariton dispersion matching that of phonon polaritons in bulk hBN, highlighting that the maximum light-matter coupling in microcavities is limited to the coupling strength between photons and the bulk material. The tunable cavity phonon polaritons could become a versatile platform for studying how the coupling strength between photons and phonons may modify the properties of polar crystals.

\section*{Introduction}

When light strongly couples to matter, new hybrid modes – polaritons - can emerge, whose coherent exchange of energy is faster than the decay rate of the original photonic modes and matter excitations \cite{mainref1,mainref2}. Recently, strong coupling (SC) between infrared light and molecular vibrations (vibrational strong coupling, VSC) has emerged as a new intriguing research topic, after it has been reported that this phenomenon can lead to modification of fundamental material properties, triggering, for example, phase transitions \cite{mainref3} or modifying electrical conductivity \cite{mainref4,mainref5} and chemical reactions \cite{mainref6,mainref7}. In the reported experiments, strong coupling was achieved by filling classical Fabry-Pérot microcavities (formed by two micrometre-scale separated mirrors) with molecules. Recently, molecular vibrational strong coupling could be achieved even on the nanometre scale, by exploiting phonon polaritons in hBN nanoresonators \cite{mainref8} and slabs \cite{mainref9}, the phonon polaritons by themselves being the results of strong coupling between infrared photons and phonons.

The strong and ultrastrong coupling between light and phonons offers intriguing possibilities for various fundamental studies and applications, including polaritonic control of THz waves in polar crystals \cite{mainref10}, which in combination with microresonators \cite{mainref11,mainref12}, can be used for the development of phonon polariton lasers \cite{mainref13,mainref14}. At surfaces or on thin layers of polar crystals, strong photon-phonon coupling can also lead to surface phonon polaritons and hyperbolic volume phonon polaritons \cite{mainref15,mainref16} that allow for nanoscale concentration of infrared and terahertz fields, which could lead to novel communication and sensing technologies \cite{mainref17,mainref18}, particularly in form of nanoresonators \cite{mainref15,mainref19,mainref20} or by coupling the polaritons with plasmonic antennas and metasurfaces \cite{mainref21,mainref22,mainref23,mainref24,mainref25}. Remarkably, a detailed study and control of the coupling strength between photons and phonons in classical Fabry-Perot microcavities is relatively unexplored terrain. This might be related with the difficulty to fabricate high-quality thin crystal layers of arbitrary thickness and place them inside the microcavities. As phonons have significant influence on many physical and chemical properties of crystals, controlling the coupling strength between infrared photons and phonons in microcavities may become an interesting platform for future fundamental and applied studies.

\begin{figure}
\centering
\includegraphics[scale=0.45]{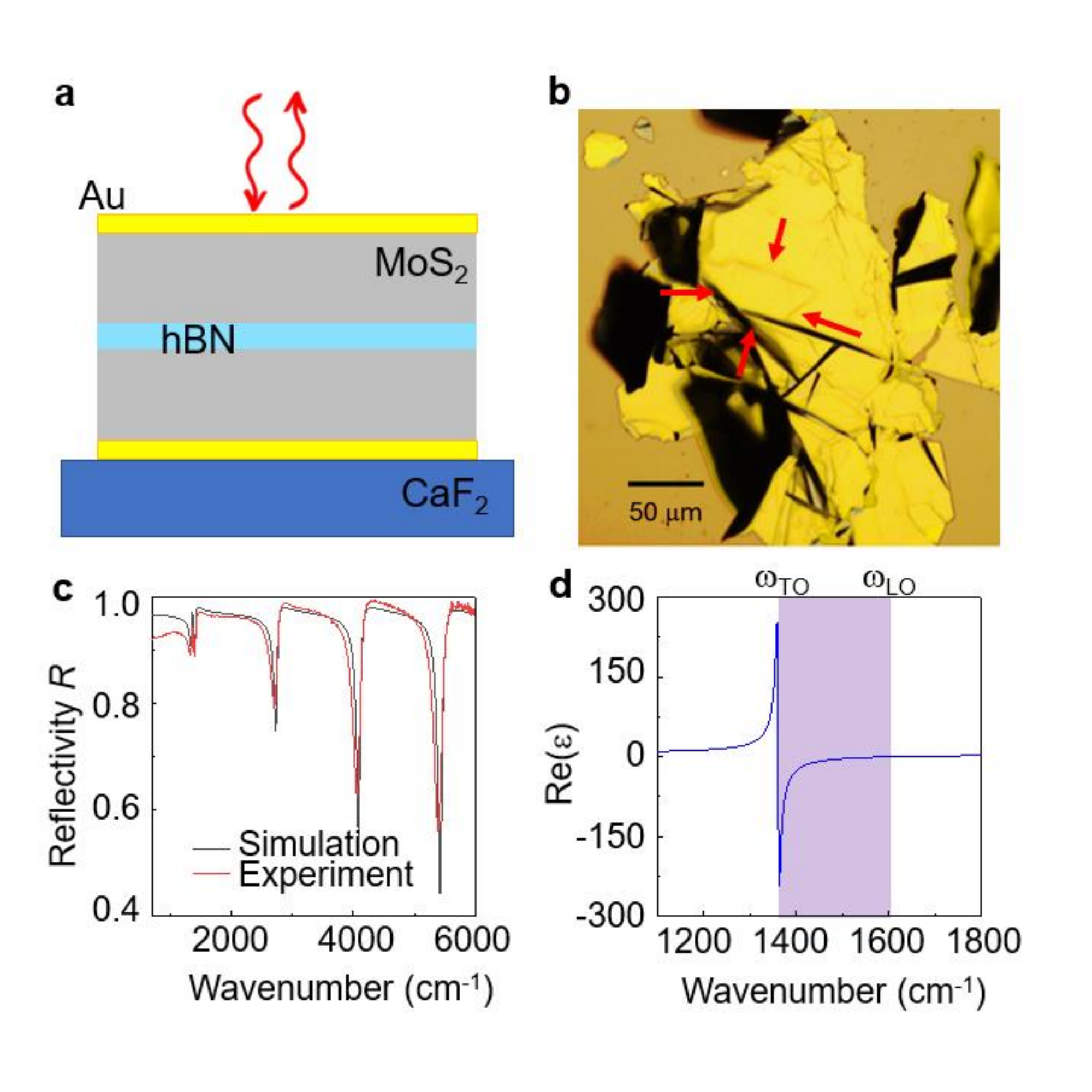}
\caption{Cavity phonon polariton experiment.  a) Sketch of the experiment, illustrating spectroscopy of the infrared light reflected at a microcavity made of two gold mirrors embedding an hBN slab. b) Optical light microscope image of a representative sample. The arrows mark the boundaries of the microcavity where the hBN flake is embedded in between two MoS2 flakes. c) Experimental (red) and simulated (black) infrared reflection spectrum of a cavity embedding a 10 nm thick hBN layer. d) Real part of the in-plane dielectric function of hBN, $\varepsilon_{hBN}$, taken from ref. \cite{mainref8}. The purple area highlights the Reststrahlen band (RB) between the transverse optical (TO) and longitudinal optical (LO) phonon frequencies.} 
\end{figure}

Here we demonstrate infrared microcavities comprising polar van der Waals (vdW) materials as a versatile test bench to study the interaction of optical phonons and photons. Importantly, the layered structure of vdW materials allows for convenient exfoliation of high-quality crystalline layers of virtually any thickness for studying the evolution of the phonon-photon coupling strengths as a function of layer thickness. Specifically, we study in this work microcavities containing high-quality layers of hexagonal Boron Nitride (hBN), which is an insulating polar material exhibiting phonons in the mid-infrared (IR) spectral range \cite{mainref26}. The experimental reflectivity of the hBN-microcavity system is well described by electrodynamical calculations based on the transfer matrix method \cite{mainref27}. Furthermore, by describing this system by two classical coupled harmonic oscillators \cite{mainref28,mainref29,mainref30}, we estimate the coupling strength between the cavity modes and the phonon excitation. We demonstrate that strong coupling can be achieved for layers as thin as a few nanometres, leading to the formation of microcavity phonon polaritons. We further systematically trace the evolution from the weak to the ultrastrong coupling (USC) regime \cite{mainref31,mainref32,mainref33,mainref34}, establishing microcavities embedding van der Waals materials as a platform for studying and tuning the coupling strengths between photons and optical phonons.

\section*{Results}

We illustrate the microcavities and the infrared reflection spectroscopy measurements in Fig. 1. The schematic in Fig. 1a represents a thin hBN flake sandwiched between two molybdenum disulphide (MoS$_2$) layers, resulting in a MoS$_2$/hBN/MoS$_2$ dielectric stack, with the hBN flake being located in the middle. Such heterostructures are fabricated following several steps of mechanical exfoliation on polydimethylsiloxane (PDMS) and deterministic transfer \cite{mainref35}. We use MoS$_2$ as a spacer, as it is spectrally flat in the mid-IR spectral region and can be easily obtained by exfoliation. An optical cavity is formed by placing the MoS$_2$/hBN/MoS$_2$ dielectric stack in between two gold layers of 20 nm thickness, fabricated by thermal evaporation. To locate the hBN flake in the cavity center - where the maximum of the electric field is expected to occur for odd cavity modes – we chose MoS$_2$ flakes of ideally the same thickness. Details of the fabrication process can be found in the Methods section. For the present study, we fabricated cavities embedding hBN flakes of varying thickness $L_{hBN}$. For each cavity, the total cavity length $L_{cav}$ was adjusted such that the fundamental cavity resonance $\omega_{cav}^{(1)}$ coincides with the frequency of the in-plane transverse optical (TO) phonon of hBN, $\omega_{TO}$ = 1364 cm$^{-1}$ (Fig. 1d) \cite{mainref15,mainref26}.  We determined $L_{cav}$ by calculating the modes of a virtual cavity (in the following referred as to a bare cavity), for which we assume a frequency-independent hBN permittivity, $\varepsilon(\omega) = \varepsilon_{hBN,\infty} = 4.52$, that neglects the phonon contribution (see Supplementary Information S1 for details).

Figure 1b shows an optical microscope image (top view) of a representative sample, where the arrows mark the microcavity containing the hBN layer. With a lateral size of typically more than 25 $\mu $m, it is large enough to perform reliable Fourier transform infrared (FTIR) microspectroscopy (illustrated in Fig. 1a). In our experiments we recorded normal-incidence reflection spectra with a FTIR setup (Bruker Hyperion 2000 infrared microscope coupled to a Bruker Vertex 70 FTIR spectrometer) that operates with a Cassegrainian objective with numerical aperture of NA = 0.4 (not shown in Fig. 1a). An example spectrum (red curve) is shown in Fig. 1c, which was obtained from a cavity embedding a 10 nm hBN layer between a 510 nm and a 370 nm thick MoS2 layer. For frequencies well above the TO phonon frequency we see a clear series of reflectivity dips that can be well matched by electrodynamical calculations employing the Transfer Matrix (TM) method (black spectrum in Fig. 1c), normal incidence. Each of the dips can be attributed to a specific cavity mode (see Supplementary Information S2). Interestingly, at the TO phonon frequency both experimental and simulated spectra reveal a double dip feature, indicating that coupling between the hBN TO phonon and the fundamental cavity mode can be achieved with our heterostructures. 

\begin{figure}
\centering
\includegraphics[scale=0.7]{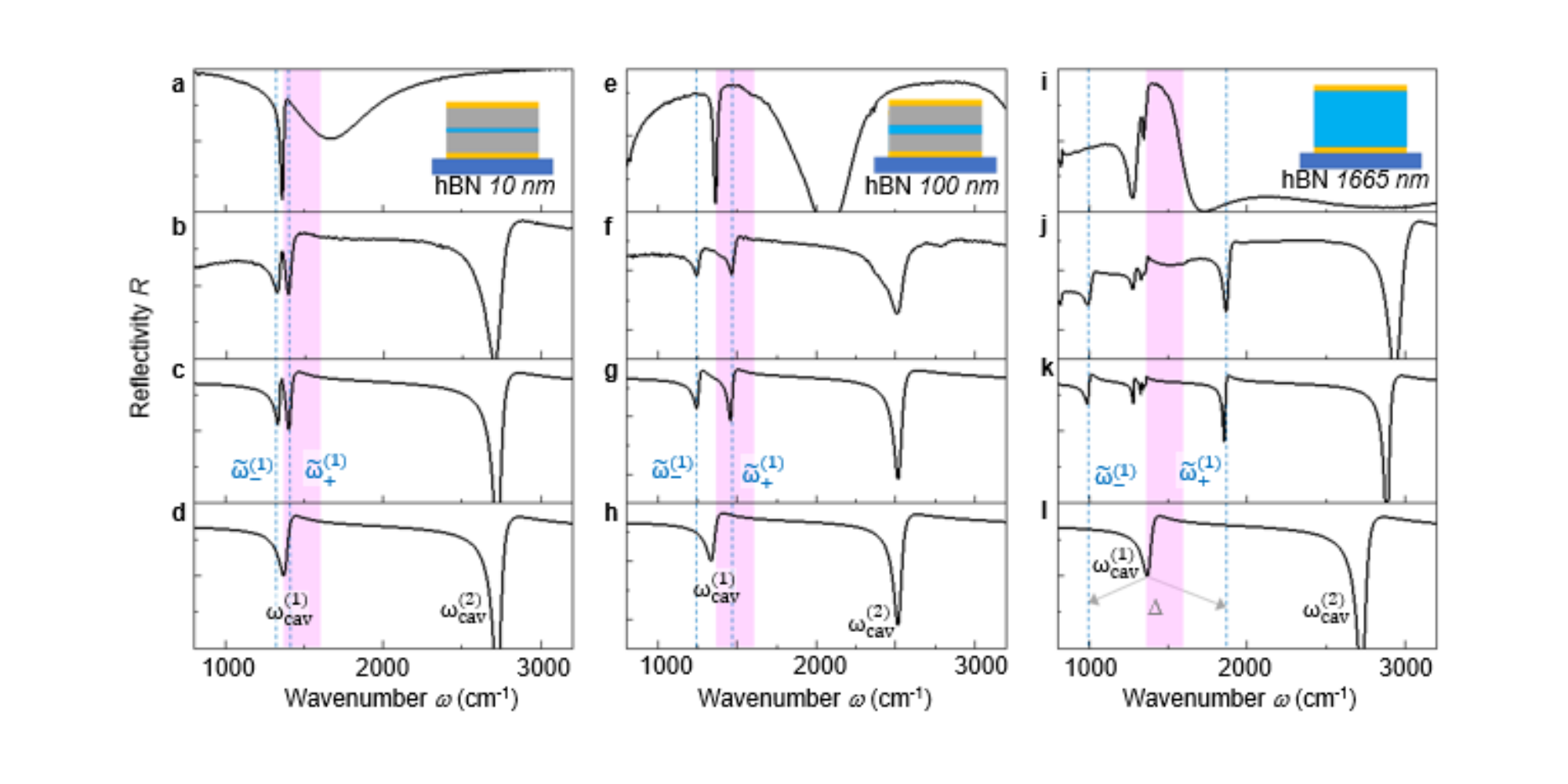}
\caption{Zoom into experimental and calculated spectra at frequencies close to the TO phonon. a) Reflectivity spectrum of the MoS$_2$/hBN/MoS$_2$ heterostructure on top of a gold mirror but without top gold mirror. Measured thicknesses are 510 nm/10 nm/370 nm. Inset illustrates the cavity cross section with top and bottom gold mirror. Reddish shaded area marks the Reststrahlen band. b) Experimental reflectivity spectrum of the stack of panel a) after closing it with top Au layer (illustrated by inset in panel a); c) Simulated reflection spectrum of cavity according to panel b), using layer thicknesses 510 nm/10 nm/370 nm. The dip splitting is marked by $\tilde{\omega}_-^{(1)}$  and $\tilde{\omega}_+^{(1)}$ . d) Simulated reflection spectrum for the cavity of panel c), in which hBN was replaced by a dielectric medium with $\varepsilon(\omega) = \varepsilon_{hBN,\infty} = 4.52$. $\omega_{cav}^{(1)}$ and $\omega_{cav}^{(2)}$  mark the first and second order cavity mode. e-h) Same as panels a-d) for a MoS$_2$/hBN/MoS$_2$ heterostructure with measured thicknesses 520 nm/100 nm/430 nm and simulated thicknesses 480 nm/100 nm/390nm. i-l) Same as panels a-d) for a completely hBN filled cavity, with measured and simulated hBN thickness 1665 nm. $\Delta = \tilde{\omega}_+^{(1)}-\tilde{\omega}_-^{(1)} $ marks the dip splitting.} 
\end{figure}

To corroborate and explore in further detail the coupling between the fundamental cavity mode and the hBN phonon, we performed a combined experimental and numerical study of cavities embedding 10 nm, 100 nm and 1665 nm thick hBN flakes (for sketches see insets in Fig. 2b,f,j). We first verify that all three hBN flakes exhibit a sharp phonon line at $\omega_{TO}$ = 1364 cm$^{-1}$, by measuring reflection spectra of the stacks prior to the fabrication of the top mirror (Figs. 2a,e,i). By closing the cavities (i.e. fabricating the top gold mirror), we clearly see a splitting of the reflection dip at the TO phonon frequency into two dips that are shifted to a lower and a higher frequency, $\tilde{\omega}_-^{(1)}$  and $\tilde{\omega}_+^{(1)}$ , respectively (Figs. 2b,f,j). Transfer matrix calculations match well the experimental spectra (Figs. 2c,g,k) upon slight modification of the nominal values of the cavity parameters (see caption of Fig. 2). The need of such modification is attributed to uncertainties in the thickness and permittivity measurements. We find that the spectral dip splitting $\Delta = \tilde{\omega}_+^{(1)}-\tilde{\omega}_-^{(1)} $ significantly increases with increasing hBN thickness, in the experiment from 76 cm$^{-1}$ to 215 cm$^{-1}$ to 864 cm$^{-1}$ (Fig. 2b,f,j) and in the calculations from  68 cm$^{-1}$, 213 cm$^{-1}$, and 860 cm$^{-1}$ (Fig. 2c,g,k). For comparison, we show in Figs. 2d,h,l transfer matrix calculations for the bare cavity, revealing that the uncoupled cavity modes indeed nearly perfectly coincide with the TO phonon frequency of the hBN layers, further corroborating that the double dip feature is a consequence of coupling between TO phonon and the respective fundamental cavity resonance.

To quantify the mode splitting and coupling strength associated to the observed splitting of the reflection dips, we theoretically analyse the evolution of the reflectivity spectra as the cavity resonance is detuned from the TO phonon. For this purpose, we performed TM simulations of the reflectivity spectra as a function of total cavity length, $L_{cav}$, while keeping fixed the thickness of the hBN layer, $L_{hBN}$ = 10 nm (Fig 3a, contour plot). By comparison with the calculated TM reflectivity spectra of the bare cavity (Fig. 3b, contour plot) we find that both the first and third bare cavity modes (recognized by the reflection dips at frequency $\tilde{\omega}_{cav}^{(1)}$ split into an upper and a lower branch of reflection dips at frequencies $\tilde{\omega}_{-}^{(1)}$ and $\tilde{\omega}_{+}^{(1)}$, featuring anti-crossing at the TO phonon frequency. The second cavity mode, in contrast, does not show any splitting of the reflection dips, since the electric field in the centre of the cavity vanishes for this mode (for further details see discussion below). The same analysis was performed with the cavity embedding the 100 nm thick hBN layer and is shown in Supplementary Information S3.

\begin{figure}
\centering
\includegraphics[scale=0.34]{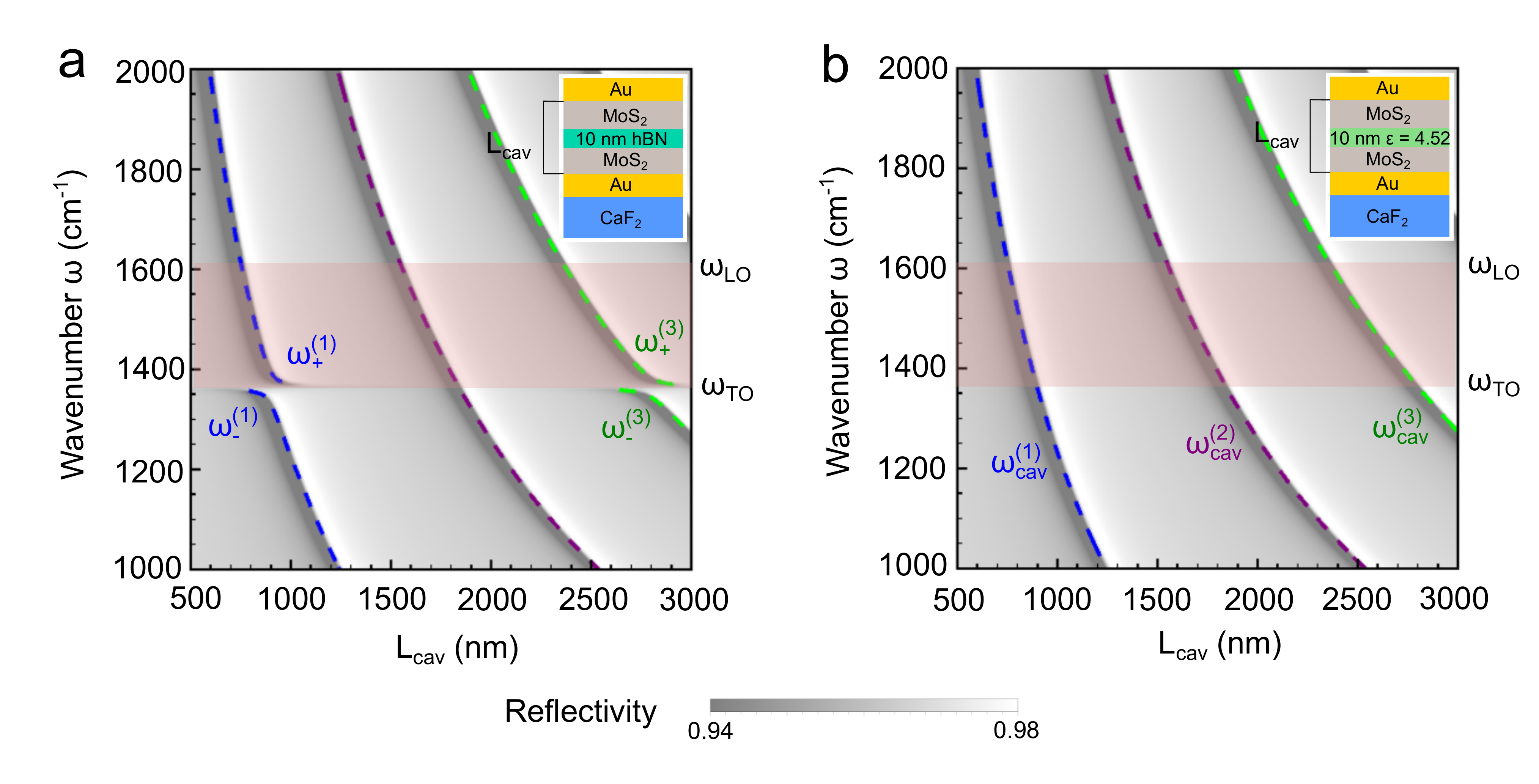}
\caption{Strong coupling in a cavity filled with 10 nm of hBN. a) TM simulated reflectivity spectra of cavity filled with 10 nm of hBN (illustrated by inset) as a function of the total cavity thickness $L_{cav}$. The dashed curves show the cavities’ eigenmode frequencies $\omega_{+}^{(j)}$and $\omega_{-}^{(1)}$ as a function of  $L_{cav}$. The reddish areas in (a) and (b) mark the upper Reststrahlen band of hBN. b) TM simulated reflectivity spectra of a bare cavity (illustrated by inset) as a function of cavity thickness  $L_{cav}$. The dashed curves show the cavities’ eigenmode frequencies $\omega_{cav}^{(j)}$ as function of $L_{cav}$.} 
\end{figure} 

We note that the spectral position of reflection minima and the actual polaritonic modes of a system can be different due to the interference of spectrally closely spaced modes (particularly in weakly coupled systems \cite{mainref36,mainref37}). For that reason, we determine the eigenfrequencies of the modes ($\omega_+^{(j)}$ and $\omega_-^{(j)}$ for the filled cavity, and $\omega_{cav}^{(j)}$ for the bare cavity), which correspond to the poles of the reflection coefficient obtained from the TM calculation \cite{mainref38} (dashed lines in Fig. 3a and b, for details see Supplementary Information S2.1 and S2.2). For $\omega_{cav}^{(1)} \approx \omega_{TO} = 1364$ cm$^{-1}$  , where the eigenmode splitting corresponds to the Rabi frequency \cite{mainref39,mainref40,mainref41,mainref42} $\Omega_R = \omega_+^{(1)}-\omega_-^{(1)}$, we find $\Omega_R$ = 67 cm$^{-1}$. From the eigenmode frequencies we can also determine the coupling strength $g$ between the hBN phonons and the microcavity photons. To that end, we model the phonon-photon interaction by the coupling of two harmonic oscillators of which one is associated with the electromagnetic field of the cavity and the other one with the hBN phonon (Supplementary Information S2.3). For small $g$ (corresponding to the weak and strong coupling regimes), the eigenfrequencies of the polaritonic modes, which correspond to the eigenvalues of the harmonic oscillator model, are given by
\begin{equation}
\omega_{\pm}^{(j)} = \frac{1}{2}(\omega_{cav}^{(j)} + \omega_{TO}) \pm \frac{1}{2} \Re\left[\sqrt{\left(\omega_{cav}^{(j)}-\omega_{TO}+i \frac{\gamma-\kappa}{2}\right)^2 + 4g^2} \right],\label{maintextEq1}
\end{equation}
where $\kappa$ and $\gamma$ are the cavity and phonon decay rates, respectively. The solutions for the eigenfrequencies for large $g$ - where the system enters the ultrastrong coupling regime – are provided in Supplementary Information S2.3, where we show that the eigenfrequencies coincide with the results obtained from the Hopfield Hamiltonian, indicating that a classical model of harmonic oscillators can describe the ultrastrong coupling regime \cite{mainref25}.

We apply Eq. \eqref{maintextEq1} to determine the coupling strength $g$ between the hBN phonon and the first-order cavity mode of Fig. 3a. To that end, we extracted $\omega_+^{(1)}$ = 1395 cm$^{-1}$,  $\omega_-^{(1)}$= 1332 cm$^{-1}$, $\kappa$ = 60 cm$^{-1}$, $\gamma$ = 5 cm$^{-1}$, and $\omega_{cav}^{(1)} = \omega_{TO}$ = 1364 cm$^{-1}$ from Fig. 3a, Fig. 1d and Fig. 2c,d. We obtain $g \approx 34$ cm$^{-1}$, which according to $\frac{g}{\kappa+\gamma} \approx 0.52 > \frac{1}{4}$ corresponds to the strong coupling regime \cite{mainref1,mainref36,mainref37}, thus revealing that a 10 nm thick layer of hBN is sufficient to achieve strong coupling between its phonons and microcavity photons. Our analysis also shows that the coupling strength is large enough to be determined in good approximation from (i) the Rabi splitting according to $\Omega_R = \omega_+^{(1)}-\omega_-^{(1)}$ = 67 cm$^{-1}$  $\approx 2g$ and even (ii) the spectral dip splitting  $\Delta = \tilde{\omega}_+^{(1)}-\tilde{\omega}_-^{(1)} = 70 $  cm$^{-1}$ $\approx 2g$ for the case that $\omega_{cav}^{(1)} \approx \omega_{TO}$.

In order to understand the evolution of the coupling strength with the hBN layer thickness, we calculate $g$ (following the procedure applied to Fig. 3) as a function of the filling factor $f = L_{hBN}/L_{cav}$ for the first and second cavity mode (blue and red solid line in Fig. 4a, respectively). We observe that $g$ increases with the filling factor for both modes, more strongly for the first mode, which is a consequence of the electric field distribution across the cavity (Fig. 4c). The second cavity mode exhibits a minimum in the center of the cavity, where the hBN layer is located. Thus, the coupling between hBN phonons and second cavity mode is much smaller than for the first cavity mode, that exhibits its field maximum in the cavity center. Note that this observation is consistent with the absence of a polariton gap for the second cavity mode in Fig. 3a. The interaction between the TO phonon and the second cavity mode remains weak until the hBN layer is thick enough to sufficiently overlap with the cavities' off-center field maxima (shown in the right panel of Fig. 4c). Interestingly, an approximate analytical expression for $g(f)$ can be obtained from a microscopic theory that is described in Supplementary Information S5.

We can identify the weak and strong coupling regimes in Fig. 4a according to the fulfilment of the conditions $\frac{g}{\kappa+\gamma} < \frac{1}{4}$ (blank area) and  $\frac{g}{\kappa+\gamma} > \frac{1}{4}$ (beige area), respectively \cite{mainref1,mainref36,mainref37}. For the first cavity mode, remarkably, the strong coupling regime starts for $f = L_{hBN}/L_{cav} \approx 0.0025$, which corresponds to hBN slabs of about 3 nm thickness (about 4 atomic h-BN layers). Moreover, the usual condition for ultrastrong coupling, $g/\omega_{TO} > 0.1$ \cite{mainref31,mainref43}  (highlighted by the green area in Fig. 4), is fulfilled for hBN layers of thickness $L_{hBN}  > 148$ nm. Interestingly, we find that  $g$ saturates for the first order mode when  $f > 0.8$ and for the second-order mode when $f > 0.9$. The maximum coupling strength is obtained for both modes when the cavity is fully filled with hBN, with $g_{max}= \sqrt{\frac{\omega_{LO}^2-\omega_{TO}^2}{4}} = 428$ cm$^{-1}$ being exclusively determined by the materials' Reststrahlen band defined by the TO and LO phonon frequencies (for further discussion see below). In this case, the ratio $g_{max}/\omega_{TO}$ reaches up to 0.31. These results show the versatility of the hBN-filled microcavities to explore the different light-matter coupling regimes, ranging from weak to ultrastrong coupling.

\begin{figure}
\centering
\includegraphics[scale=0.35]{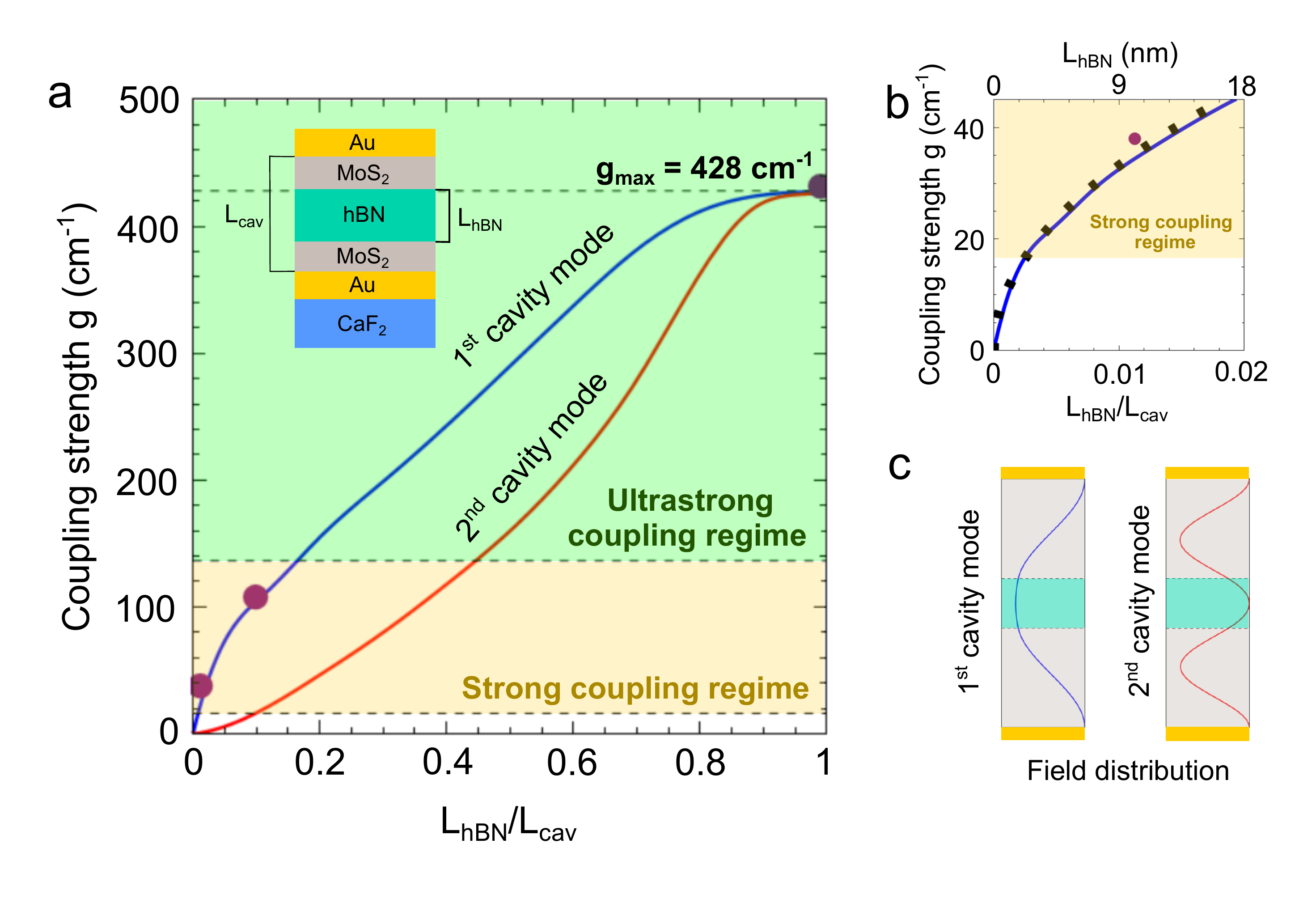}
\caption{Evolution of the coupling strength as a function of the cavity filling factor. a) Coupling strength $g$ between the TO phonon and the bare-cavity modes, for a hBN layer of thickness $L_{hBN}$ in the middle of a cavity of thickness $L_{cav}$, as sketched in the inset. The calculated evolution of $g$ for varying $L_{hBN}$ is shown for the first (blue solid line) and the second (red solid line) cavity modes. The coupling strengths corresponding to the strong coupling regime are highlighted by the beige area, and the green area corresponds to the ultrastrong coupling regime. Purple symbols indicate the experimental coupling strength $g_{exp}$ obtained from Fig. 2.  b) Zoom into panel (a) for small filling factors $L_{hBN}/L_{cav}$, showing the calculated $g$  for the first cavity mode. The analytical approximation $g\approx 332\sqrt{L_{hBN}/L_{cav}}$ cm$^{-1}$ is shown by black dots. c) Sketch of the amplitude of the electric field distribution of the first and second cavity modes.} 
\end{figure} 

Figure 4b shows a zoom into Fig. 4a for small filling factors (when the field distribution can be assumed homogeneous within the hBN), where we observe that the evolution of the coupling strength of the first cavity mode scales with $\sqrt{L_{hBN}}$   (solid dots). This behavior is consistent with the well-known scaling law of strong coupling \cite{mainref44,mainref45,mainref46}, $g\propto \sqrt{N}$, where $N$ is the number of oscillators, which in case of a hBN slab is proportional to $L_{hBN}$ (see Supplementary Information S5.2 for the derivation of a simplified analytical expression for very small and very large filling factors).  

To determine the coupling regime for the experimental spectra shown in Fig. 2, we determine the coupling strength according to $g_{exp} = \Delta/2$, where $\Delta = \tilde{\omega}_+^{(1)}-\tilde{\omega}_-^{(1)}$ is the spectral splitting of the reflectivity dips (indicated in Fig. 2l). This approach \cite{mainref36,mainref47} is valid only for large $g$ and when $\omega_{cav}^{(1)} \approx \omega_{TO}$, both conditions being fulfilled in our experiments (see Fig. 3 and corresponding discussion above). Plotting $g_{exp}$ in Fig. 4a (purple symbols), good agreement with the theoretical values (blue solid line) is found, verifying that strong coupling between microcavity photons and phonons can be indeed achieved experimentally with layers of a polar material as thin as 10 nm. 

\begin{figure}
\centering
\includegraphics[scale=0.33]{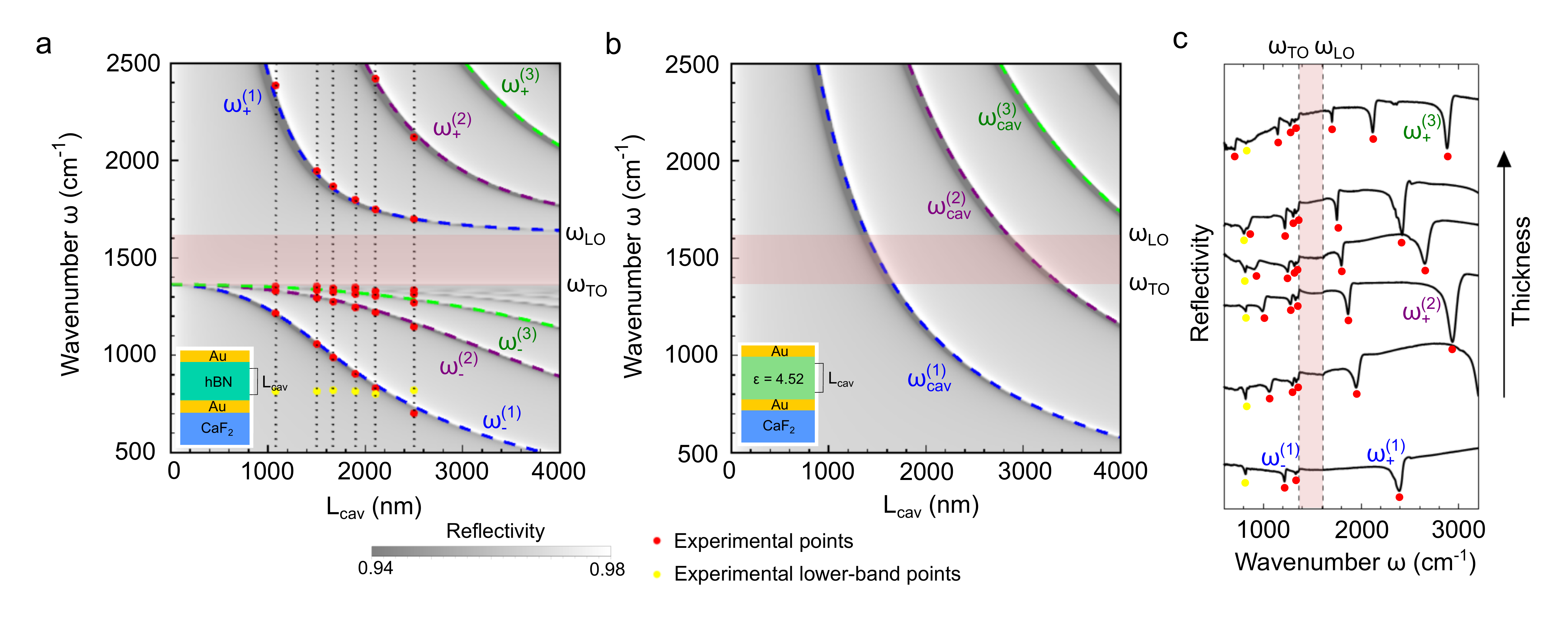}
\caption{Ultrastrong coupling in cavities fully filled with hBN. a) Reflectivity spectra of a cavity fully filled with hBN (illustrated by inset) as a function of cavity thickness $L_{cav}$. The dashed curves show the cavities' eigenmode frequencies $\omega_{+}^{(1)}$ and $\omega_{-}^{(1)}$ as a function of $L_{cav}$. Red and yellow symbols show the spectral position of the experimental reflectivity dips extracted from panel (c). The reddish areas in (a) and (b) mark the upper Reststrahlen band of hBN; b) Reflectivity spectra of a bare cavity (illustrated by inset) as a function of cavity thickness $L_{cav}$. The dashed curves show the cavities' eigenmode frequencies $\omega_{cav}^{(1)}$ as function of $L_{cav}$; c) Experimental reflectivity spectra of cavities fully filled with hBN (vertically offseted). Cavity thicknesses (from bottom to top) are: 1080 nm, 1500 nm, 1665 nm, 1900 nm, 2100 nm, 2500 nm.} 
\end{figure} 

To further illustrate and explore the regime of maximum coupling strength, we performed a systematic experimental and theoretical study of cavities fully filled with hBN (Fig. 5a). Analogously to Fig. 3, we calculated reflection spectra (contour plots) and eigenmode frequencies (dashed lines) as a function of the cavity (i.e. hBN) thickness $L_{hBN}=L_{cav}$. By comparison with the spectra and eigenmodes of the corresponding bare cavities (Fig. 5b), we clearly recognize the anti-crossing between the hBN phonon ($\omega_{TO}$) and all cavity modes \cite{mainref48}  ($\omega_{cav}^{(j)}  $ with $j$ = 1, 2, 3... ), yielding the polaritonic eigenmodes $\omega_{+}^{(j)}$ and $\omega_{-}^{(j)}$ (contrary to Fig. 3, where coupling is observed only for odd $j$). Further, a much larger spectral separation of the upper and lower polariton branches occurs as compared to Fig. 3a ($L_{hBN}$= 10 nm), yielding a polaritonic gap spanning the whole Reststrahlen band (reddish area in Fig. 5a,b). We determine a Rabi splitting of $\Omega_R \approx 2g$ = 856 cm$^{-1}$, which lies well inside the ultrastrong coupling regime ($g/\omega_{TO}= 0.31$), and which is larger than the Reststrahlen band (246 cm$^{-1}$). We experimentally confirm the calculations by measuring reflectivity spectra of differently thick cavities that are fully filled with hBN (Fig. 5c). The spectral positions of the reflectivity dips are plotted in Fig. 5a (red symbols), showing an excellent agreement with the calculated spectra and cavity eigenmodes. We note that the experimental reflectivity spectra exhibit a set of thickness-independent dips at around 819 cm$^{-1}$ (yellow dots in Fig. 5b), which stem from the out-of-plane phonon of hBN \cite{mainref32}. This phonon appears in the experimental spectra due to the focused illumination of the cavity using a Cassegrainian objective, yielding electric field components perpendicular to the hBN layer that can couple to the out-of-plane hBN phonon (see Supplementary Information S4).

\begin{figure}
\centering
\includegraphics[scale=0.7]{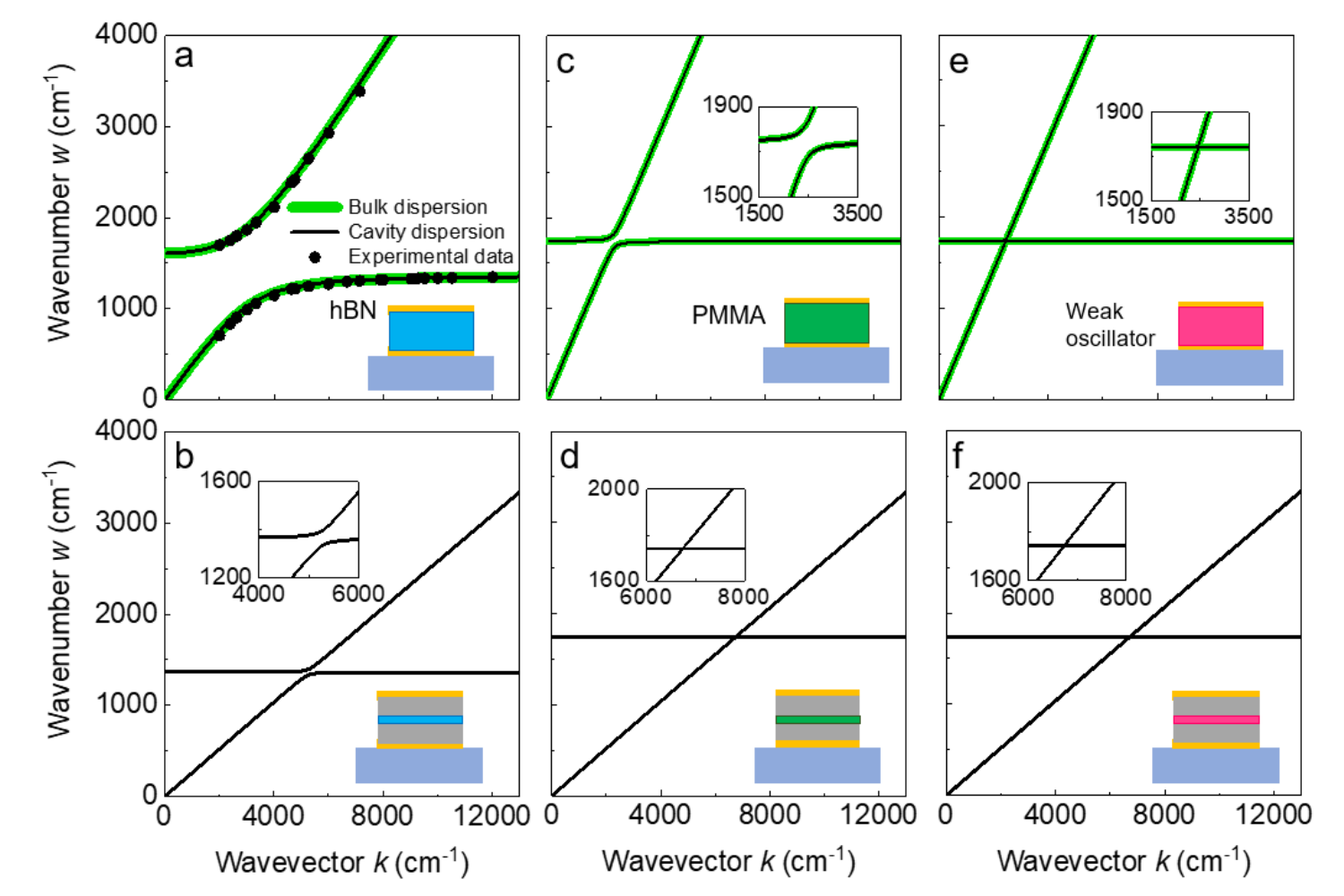}
\caption{Cavity polariton dispersions. a) Measured (black symbols) and calculated (black curves) cavity phonon polariton dispersion for a cavity fully filled with hBN. Green curves show the bulk phonon polariton dispersion of hBN obtained with Eq. 2. b) Calculated (black curves) cavity phonon polariton dispersion of the first mode of a cavity filled with a 10 nm thick hBN layer. c) Calculated (black curves) cavity polariton dispersion for a cavity fully filled with molecules exhibiting a C=O vibration. Green curves show the bulk polariton dispersion of the filling material. d) Dispersions for a 10 nm thick molecular layer embedded in the cavity. e) Calculated (black curves) cavity polariton dispersion for a cavity fully filled with molecules exhibiting a vibration of reduced oscillator strength. Green curves show the bulk polariton dispersion of the filling material. f) Dispersions for a 10 nm thick molecular layer of reduced coupling strength embedded in the cavity. Schematics illustrate the cavity cross sections.} 
\end{figure}

To determine and compare the dispersions of the microcavity phonon polaritons, $\omega(k)$, we extracted the wavevector k from the reflectivity spectra according to
\begin{equation}
k = \frac{j \pi}{L_{cav}},
\end{equation}
assuming perfect metal mirrors and where $L_{cav}$ is the cavity length at which the eigenmode of order $j$ and frequency $\omega$ is found. In Fig. 6a,b we compare $\omega(k)$ for the cavity that is fully filled with hBN (obtained from Fig. 5) and the cavity embedding a 10 nm thick hBN layer (obtained from Fig. 3), respectively (calculations are shown by black curves and experimental values by black symbols). Note that for the fully filled cavity, all modes $j$ were considered, whereas for the cavity filled with 10 nm of hBN we considered only the first mode ($j= 1$). Both dispersions feature anti-crossing, separating into a lower and an upper microcavity phonon polariton branch, with large Rabi splitting amounting to $\Omega_R \approx 856$ cm$^{-1}$ and $\Omega_R \approx 67$ cm$^{-1}$, respectively. To appreciate the dramatic coupling strengths between infrared microcavity modes and phonons with respect to typical molecular vibrations \cite{mainref34}, we show in Figs. 6c,d the mode dispersions obtained (analogous to Fig. 3 and 5) for cavities embedding molecules that possess C=O vibrations. Specifically, we consider the ensemble of C=O oscillators of poly(methyl methacrylate) (PMMA), whose electromagnetic response is described by the dielectric function provided in the Supplementary Information S1. Clear anti-crossing can be observed for the cavity fully field with PMMA (Fig. 6c), but the Rabi splitting $\Omega_R \approx 159$ cm$^{-1}$ is more than 5 times smaller than that of the cavity fully filled with hBN. For the cavity filled with a 10 nm thick PMMA layer we do not find anti-crossing (Fig. 6d), revealing that the system is in the weak coupling regime, in contrast to the strong coupling regime achieved with a 10 nm thick hBN layer (Fig. 6b). We note that the C=O vibrations are rather strong and that many molecular vibrations can be much weaker. To demonstrate the coupling between microcavity modes and weaker molecular oscillators by way of an example, we reduced the oscillator strength in the permittivity model of C=O by a factor of 100 and recalculated the dispersions. In both the partially and fully filled cavity we do not observe anti-crossing (Fig. 6e,f), highlighting that for weak molecular oscillator strong coupling cannot be achieved by placing them into a microcavity.

We finally compare the dispersions of the fully filled cavities with the bulk dispersion of the corresponding filling material, $\omega(k) = \frac{ck}{\sqrt{\varepsilon_m}}$,  where $c$ is the speed of light and $\varepsilon_m$ the dielectric function of the filling material (green solid lines in Fig. 6a,c and e). Interestingly, we find that the cavity mode dispersion is identical to that of the bulk dispersion of the filling material \cite{mainref49}, independent of whether phonon or molecule oscillators (weak or strong) are embedded into the cavity. The maximum splitting is determined exclusively by the material properties as clearly shown by the analytical expression of $g$ in the Supplementary Information S5 and is also highlighted in ref. \cite{mainref50}. These results show that the maximum coupling strength between a cavity mode and a dipolar excitation is governed by that of photons and bulk, implying that fully filling a resonant cavity with a specific material does not enhance the coupling strength between light and matter. The cavity merely enforces the strongly coupled state by selecting the corresponding wavevector. Importantly, strong coupling can be only achieved in case the bulk dispersion of the cavity's filling material exhibits strong coupling, i.e. anti-crossing. In case the dispersion of light in bulk does not exhibit strong coupling (i.e. no polaritonic behavior), the coupling of this bulk material with a cavity mode will remain weak, that is, the cavity does not leverage the coupling strength beyond the one that can be achieved with bulk. Interestingly, it has been reported that vibrational strong coupling in cavities can modify physical and chemical properties of the cavity filling \cite{mainref2,mainref51,mainref52}, although the coupling strength is not enhanced. An effect other than the coupling strength - such as a modification of the density of states - may be needed to explain this intriguing phenomenon.

In conclusion, we demonstrated that classical microcavities can be applied for studying and tuning the coupling between photons and optical phonons in thin layers of polar vdW materials, which can be obtained in high crystal quality by exfoliation. For the studied vdW material hBN, our theoretical analysis predicts strong coupling for hBN layers between 3 and 148 nm thickness, and ultrastrong coupling for hBN thicknesses larger than 148 nm. Analysis of experimental reflection spectra of cavities embedding 10 and 100 nm-thick hBN layers confirms strong coupling. For fully filled cavities ultrastrong was demonstrated experimentally. In comparison with typical molecular vibrational strong coupling with fully filled cavities, the coupling strength is about five times larger. Our experiments can be readily adapted to study phonons in other vdW materials \cite{mainref53}, including doped semiconductors and heterostructures exhibiting multiple phonons or plasmon-phonon coupling. Considering the intriguing phenomena that have been observed due to molecular vibrational strong coupling, it may be interesting to study how ultrastrong coupling of phonons may affect physical and chemical properties of materials that are embedded into cavities. Only recently it has been predicted that strong coupling in quantum paraelectrics can trigger the ferroelectric phase \cite{mainref54}.

\section*{Methods}

\subsection*{Fabrication of fully filled microcavities}

First, the bottom optical mirror was prepared on a CaF$_2$ substrate by thermally evaporating 20 nm of gold. The evaporation was performed in high vacuum conditions (pressure $10^{-6}$ mbar) at a rate of 0.8 nm/min.

Large thick flakes of hexagonal Boron Nitride (hBN) were obtained via mechanical exfoliation of commercially available hBN crystals (HQ Graphene Co.) using blue Nitto tape (Nitto SPV 224P). Subsequently, the blue Nitto tape was placed on a polydimethylsiloxane (Gelpack PF GEL film WF 4, 17 mil.) transparent stamp. This way, after removing the blue Nitto tape from the polydimethylsiloxane stamp, some flakes remain on the polydimethylsiloxane stamp. Flakes of approximately the desired thickness were optically identified. The chosen ones were transferred on top of the bottom optical mirror using the deterministic dry transfer technique \cite{mainref35}. To have a better characterization of the thickness, the hBN flakes on top of the bottom optical mirror were measured using a profilometer (Dektak 150). Large areas of the aimed thickness (ranging from 1000 to 2500 nm) were consequently identified and localized.

Finally, the sample was covered again with 20 nm of thermally evaporated gold, to form the top optical mirror. This evaporation was performed under the same conditions as for the one conforming the bottom optical mirror.

\subsection*{Fabrication of microcavities embedding thin layers of hBN}

Using the same techniques as for the case of hBN flakes, flakes of MoS$_2$ were exfoliated from MoS$_2$ crystals (SPI supplies). The flakes were characterized and transferred onto a bottom optical mirror, prepared as in the previous sample (20 nm of gold thermally evaporated on a CaF$_2$ substrate). Thin hBN flakes were obtained and optically identified, then transferred on the MoS$_2$ flake already on the gold layer. The heterostructure thickness was then characterized with the profilometer. A second MoS$_2$ flake of ideally the same thickness as the first one was transferred on top of the MoS$_2$-hBN structure. In order to obtain the final thickness of the stack, it was once again characterized with the profilometer. A second layer of gold 20 nm thick was then thermally evaporated over the structure.

\section*{Acknowledgements}

This work is supported by the Spanish Ministry of Science and Innovation under the María de Maeztu Units of Excellence Program (MDM-2016-0618), and Projects PID2019-107432GB-I00 and RTI2018-094861-B-100; by the European Union H2020 under the Marie Curie Actions (796817-ARTEMIS); and by Project PI2017-30 and Grant IT1164-19 for research groups of the Basque University system from the Department of Education of the Basque Government.

\cfoot {\thepage}

\newpage
\clearpage
\renewcommand{\theequation}{S\arabic{equation}}
\renewcommand{\thesection}{S\arabic{section}}
\renewcommand{\thefigure}{S\arabic{figure}}

\renewcommand{\thepage}{S\arabic{page}}  
\setcounter{page}{1}
\setcounter{equation}{0}
\setcounter{figure}{0}
\setcounter{section}{0}

\begin{center}
\begin{LARGE}

\textbf{Supplementary Information for:}
\medskip

\textbf{Microcavity phonon polaritons – from weak to ultra-strong phonon-photon coupling}
\end{LARGE}

\bigskip

\begin{large}
María Barra-Burillo$^{1}$, Unai Muniain$^{2}$, Sara Catalano$^{1}$, Marta Autore$^{1}$, Felix Casanova$^{1,3}$, Luis E. Hueso$^{1,3}$, Javier Aizpurua$^{2,4}$, Ruben Esteban$^{2,4}$ and Rainer Hillenbrand$^{3,5}$ 
\end{large}
\medskip

\begin{small}
$^1$\textit{CIC nanoGUNE BRTA, 20018 Donostia-San Sebastián, Spain}

$^2$\textit{Donostia International Physics Center, 20018 Donostia-San Sebastián, Spain}

$^3$\textit{IKERBASQUE, Basque Foundation for Science, 45011 Bilbao, Spain}

$^4$\textit{Materials Physics Center, CSIC-UPV/EHU, 20018 Donostia-San Sebastián, Spain}

$^5$\textit{CIC nanoGUNE BRTA and EHU/UPV, 20018 Donostia-San Sebastián, Spain }

\end{small}
\end{center}

\section{Description of the cavities and permittivities of the different materials} \label{sec_description}

\begin{figure}
\centering
\includegraphics[scale=0.85]{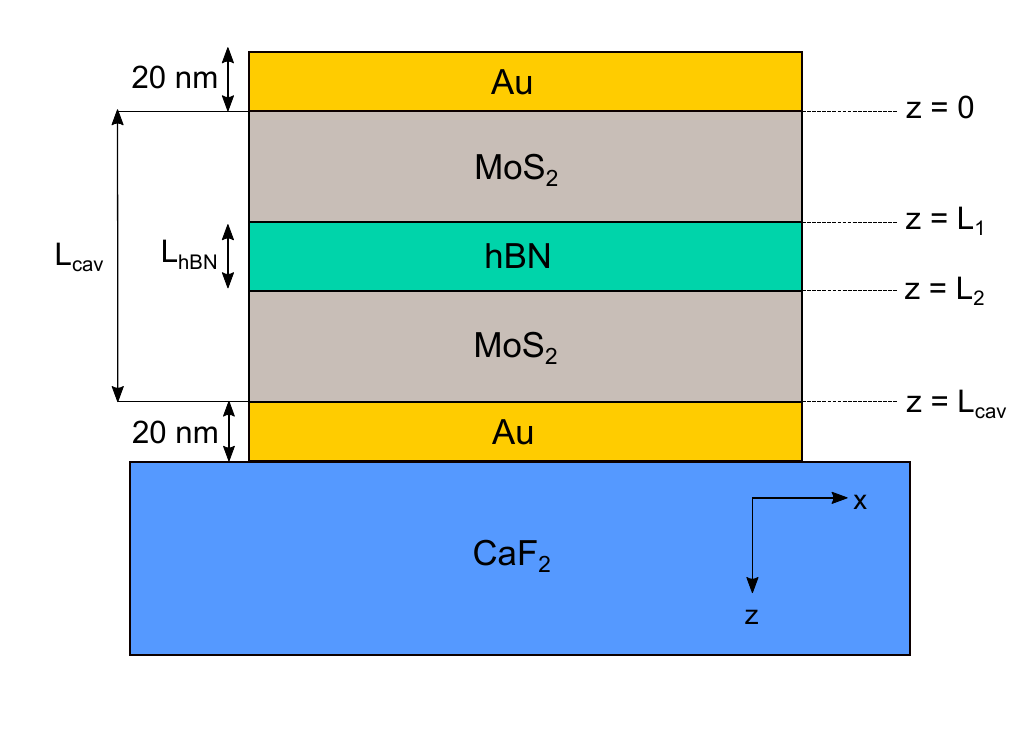}
\caption{Schematic diagram of the hBN microcavities. We indicate the coordinates axes and the material and thickness of each layer. In some simulations, hBN is substituted by PMMA, a weak oscillator or a material with constant permittivity $\varepsilon_{hBN,\infty}$.} \label{figure_cavity_diagram}
\end{figure}

In this section, we describe in more detail the hexagonal boron nitride (hBN) microcavities that we analyze in this work, including all the different thicknesses. We use Fabry-Pérot cavities containing  hBN layers of variable thickness. A schematic diagram is shown in Fig. \ref{figure_cavity_diagram}. The cavities are formed by planar layers, and all the interfaces are perpendicular to the $z$ axis, as shown by the coordinates axes included in the scheme. In all the calculations the incident medium (i.e. the medium from which the system is illuminated) is vacuum, the substrate is CaF$_2$ and the mirrors are 20 nm-thick gold layers. The inside of the cavity, i.e. without considering the mirrors, extends from $z=0$ to $z=L_{cav}$  ($L_{cav}$ is the total thickness), and it contains a layer of thickness $L_{hBN}$ placed between $z=L_1$ and $z=L_2$. In all the experiments and in many of the transfer-matrix simulations, the material of this layer is hBN, and we refer to this particular case as a \textit{hBN-filled cavity}. In other simulations, we replace hBN by a material with constant permittivity $\varepsilon_{hBN,\infty}$, which corresponds to the high-frequency permittivity of hBN (see below). We refer to these cavities as \textit{bare cavities}, because the contribution of the hBN phonons is eliminated. The rest of the cavity (the layers between $z=0$ and $z=L_1$ and between $z=L_2$ and $z=L_{cav}$) is filled by MoS$_2$, which does not show any resonant feature in the analyzed range of frequencies. 

In the transfer-matrix simulations, we have modelled the materials of the system as follows. For the incident medium we use the vacuum permittivity, and the relative permittivity of the substrate material CaF$_2$ is $\varepsilon_{CaF_2} = 1.882$. The gold mirrors are described by a Drude function that fits the low-frequency experimental data from Ref. \cite{johnson72}:
\begin{equation}
\varepsilon_{Au}(\omega) = 1-\frac{\omega_{p,Au}^2}{\omega(\omega + i\gamma_{Au})},
\end{equation}
with plasma frequency $\omega_{p,Au} = 73114.15 \text{ cm}^{-1}$ and damping frequency $\gamma_{Au} = 571.04 \text{ cm}^{-1}$ \cite{esteban15}.

We model hBN with a diagonal permittivity tensor $ \overset\leftrightarrow{\varepsilon}_{hBN}(\omega)$, in order to take into account the anisotropy of this material. hBN is characterized by a layered atomic structure, so that the phonons that are polarized along the plane of the atomic layers (in-plane phonons) have different frequencies compared to those polarized in the normal direction, i.e. parallel to the anisotropic axis ($\parallel$, out-of-plane phonons). In our configuration, the plane of the atomic layers is parallel to all the Fabry-Pérot interfaces ($x-y$ plane). We thus have $ \overset\leftrightarrow{\varepsilon}_{hBN}(\omega) = \text{diag}\left(\varepsilon_{hBN}(\omega),\varepsilon_{hBN}(\omega),\varepsilon_{hBN,\parallel}(\omega)\right)$, where we omit the second subscript in the $x$ and $y$ components for brevity (notice that the $x$ and $y$ components are identical). In particular, the in-plane permittivity tensor component  $\varepsilon_{hBN}(\omega)$ follows a Lorentzian function  \cite{autore18}:
\begin{equation}
\varepsilon_{hBN}(\omega) = \varepsilon_{hBN,\infty} \left( 1 - \frac{\omega_{LO}^2-\omega_{TO}^2}{\omega_{TO}^2-\omega^2-i\omega \gamma} \right),\label{hBN_lorentzian}
\end{equation}
where $\omega_{TO}$ is the corresponding transverse optical (TO) phonon frequency, $\omega_{LO}$ is the longitudinal optical (LO) phonon frequency, $\gamma$ is the damping frequency and $\varepsilon_{hBN,\infty}$ is the high-frequency permittivity. For our analysis, the exact value of $\omega_{TO}$ determines the detuning between the cavity mode and the phonon, which is a crucial parameter in strong coupling. Therefore, to obtain the value of this parameter in our samples, we measured the reflectivity spectra of a 100 nm-thick hBN layer before the fabrication of the cavities. A clear resonance was observed at $\omega_{TO} = 1364 \text{ cm}^{-1}$, which is the value that we consider in the simulations. This frequency is very close to the value $\omega_{TO} = 1360 \text{ cm}^{-1}$ reported in Ref. \cite{autore18}. For all the other parameters in Eq. \eqref{hBN_lorentzian} we use the values of Ref. \cite{autore18}, $\varepsilon_{hBN,\infty} = 4.52$, $\omega_{LO} = 1610 \text{ cm}^{-1}$ and $\gamma = 5 \text{ cm}^{-1}$.

Equation \eqref{hBN_lorentzian} properly describes the optical response of hBN in all the results in the main text, because the electric fields are directed along the $x$ direction and thus hBN can be treated as an isotropic material with the permittivity given by Eq. \eqref{hBN_lorentzian}. This approach is also followed in the S.I. when not stated otherwise. However, in Sec. \ref{section_focused_beam} we consider focused illumination, where it is necessary to take the anisotropy of hBN into account. In this case, the permittivity tensor component along the $z$ axis $\varepsilon_{hBN,\parallel}(\omega)$ follows the same Lorentzian function as in Eq. \eqref{hBN_lorentzian}, except that the values of the parameters are  $\varepsilon_{hBN,\infty} = 4.52$, $\omega_{TO,\parallel} = 746 \text{ cm}^{-1}$, $\omega_{LO,\parallel} = 819 \text{ cm}^{-1}$ and $\gamma_\parallel = 4 \text{ cm}^{-1}$ (taken from Ref. \cite{autore18}).

\begin{figure}
\centering
\includegraphics[scale=0.5]{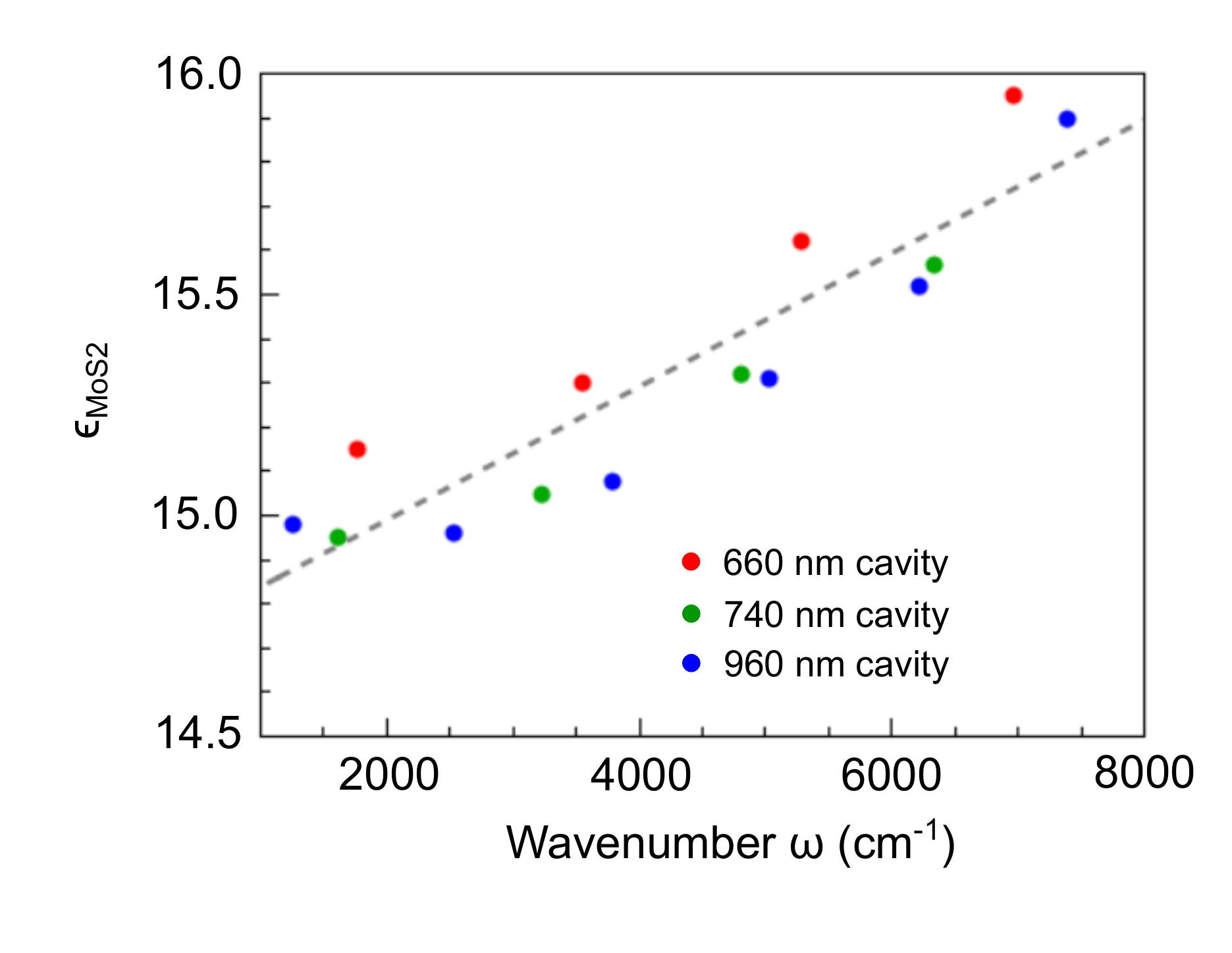}
\caption{Linear fitting of the permittivity of MoS$_2$. The dots indicate the experimental resonant wavenumber of the different Fabry-Pérot modes  and the corresponding permittivity extracted from these values according to the procedure described in the text, for three cavities fully filled by MoS$_2$ of nominal thickness $L_{cav}$ = 660 nm (red dots), $L_{cav}$ = 740 nm (green dots) and $L_{cav}$ = 960 nm (blue dots). The dashed line corresponds to the linear fitting of all dots, given by Eq. \eqref{permittivity_mos2}.} \label{interpolation_mos2}
\end{figure}

In order to obtain the permittivity of MoS$_2$, we have fabricated three  microcavities fully filled by MoS$_2$ of different thickness: $L_{cav}$ = 660 nm, 740 nm and 960 nm. The geometry is the same as in Fig. \ref{figure_cavity_diagram} but without the hBN layer. We measure the reflectivity spectra and obtain the frequencies of the Fabry-Pérot modes for each cavity up to 8000 cm$^{-1}$. We then perform transfer-matrix simulations of the cavities and find the value of the permittivity $\varepsilon_{MoS_2}$ required to reproduce the position of the experimental dips. We show in Fig. \ref{interpolation_mos2} the frequencies of the dips for the three cavities and their corresponding value of $\varepsilon_{MoS_2}$. A linear fitting of these results (dashed line) gives the permittivity that we use in the simulations:
\begin{equation}
\varepsilon_{MoS_2}(\omega) = 14.689 + 0.000151\omega,\label{permittivity_mos2}
\end{equation}
where $\omega$ is in units of cm$^{-1}$.

Last, in some simulations in the end of the main text, we replace hBN with PMMA and a weak oscillator. For PMMA, we use a permittivity that also follows a Lorentzian form \cite{menghrajani19}:
\begin{equation}
\varepsilon_{PMMA}(\omega) = \varepsilon_{PMMA,\infty} + F_{PMMA}\frac{\omega_{PMMA}^2 }{\omega_{PMMA}^2-\omega^2-i\omega \gamma_{PMMA}},
\end{equation}
with high-frequency permittivity $\varepsilon_{PMMA,\infty} = 1.99 $, vibrational frequency $\omega_{PMMA} = 1742 \text{ cm}^{-1}$, damping frequency $\gamma_{PMMA} = 13 \text{ cm}^{-1}$ and oscillator strength $F_{PMMA} = 0.0165$. When we consider a weak oscillator in Figs. 6e and 6f in the main text, we use the same values as for PMMA, except for  $F_{PMMA} = 1.65 \cdot 10^{-4}$.

\section{Characterization of the polaritonic frequencies}

\subsection{Transfer-matrix simulations}\label{subsec_transfer_matrix}

We first summarize briefly how to calculate the reflectivity spectra of hBN microcavities using the transfer-matrix formalism \cite{passler17}. This method allows to obtain the total Fresnel reflection $r_{total}^{s(p)}$  and transmission $t_{total}^{s(p)}$  coefficients of a $s$($p$)-polarized planewave, for any layered structure, in terms of the thickness $d_i$ and permittivity $\varepsilon_i$ of each layer $i$ and the Fresnel coefficients $r^{s(p)}_{i,i+1}$  and $t^{s(p)}_{i,i+1}$  of each interface between layers $i$ and $i+1$. In total, the system contains $N_{lay}$ finite-sized layers, and we label the incident medium as $i=0$ and the substrate as $i=N_{lay}+1$. For the calculations in the main text, we consider that the light is incident along the $z$ direction perpendicular to the planar interfaces and all materials are considered isotropic. However, for the calculations in Sec. \ref{section_focused_beam}, we analyze the effects of the anisotropy of hBN and, thus, here we discuss the complete formalism for anisotropic materials. To take the anisotropy of hBN into account, we consider that each layer $i$ has a permittivity tensor of the general diagonal form $ \overset\leftrightarrow{\varepsilon_i} = \text{diag}(\varepsilon_{i,x},\varepsilon_{i,y},\varepsilon_{i,z})$ ($\varepsilon_{hBN,x} = \varepsilon_{hBN,y}$ for hBN in our experiments). Light is incident in the $x-z$ plane.  The Fresnel coefficients are given by \cite{passler17}
\begin{align}
&r_{i,i+1}^s = \frac{\sqrt{\varepsilon_{i,y}-\xi^2}-\sqrt{\varepsilon_{i+1,y}-\xi^2}}{\sqrt{\varepsilon_{i,y}-\xi^2}+\sqrt{\varepsilon_{i+1,y}-\xi^2}}, \qquad 
&r_{i,i+1}^p = \frac{\varepsilon_{i+1,x}\sqrt{\varepsilon_{i,x}\left(1-\frac{\xi^2}{\varepsilon_{i,z}}\right)}-\varepsilon_{i,x}\sqrt{\varepsilon_{i+1,x}\left(1-\frac{\xi^2}{\varepsilon_{i+1,z}}\right)}}{\varepsilon_{i+1,x}\sqrt{\varepsilon_{i,x}\left(1-\frac{\xi^2}{\varepsilon_{i,z}}\right)}+\varepsilon_{i,x}\sqrt{\varepsilon_{i+1,x}\left(1-\frac{\xi^2}{\varepsilon_{i+1,z}}\right)}},\nonumber \\
&t_{i,i+1}^s = \frac{2\sqrt{\varepsilon_{i,y}-\xi^2}}{\sqrt{\varepsilon_{i,y}-\xi^2}+\sqrt{\varepsilon_{i+1,y}-\xi^2}}, \qquad 
&t_{i,i+1}^p = \frac{2\varepsilon_{i+1,x}\sqrt{\varepsilon_{i,x}\left(1-\frac{\xi^2}{\varepsilon_{i,z}}\right)}}{\varepsilon_{i+1,x}\sqrt{\varepsilon_{i,x}\left(1-\frac{\xi^2}{\varepsilon_{i,z}}\right)}+\varepsilon_{i,x}\sqrt{\varepsilon_{i+1,x}\left(1-\frac{\xi^2}{\varepsilon_{i+1,z}}\right)}},
\end{align}
where $\xi = k_x / k_0$ is the parallel component of the wavevector $k_x$ normalized with respect to the wavevector in vacuum $k_0$.

The total transfer matrix $\mathbf{T}^{s(p)}$ relates the amplitudes of the electric field in the incident medium ($i=0$) and the substrate ($i=N_{lay}+1$) for $s(p)$-polarized light: $\begin{pmatrix}
E_{0,+}^{s(p)}\\
E_{0,-}^{s(p)}
\end{pmatrix} = \mathbf{T}^{s(p)} \begin{pmatrix}
E_{N_{lay}+1,+}^{s(p)}\\
E_{N_{lay}+1,-}^{s(p)}
\end{pmatrix}$. The second subindex indicates the direction of propagation. + corresponds to the direction of the incoming planewave (and thus also to the direction of the transmitted light) and, similarly, the subindex - corresponds to the direction of the reflected light. In our system, $E_{N_{lay}+1,-}^{s(p)} = 0$, because in the substrate the electric field only propagates in the direction of the transmission. $\mathbf{T}^{s(p)}$ is given by the expression
\begin{equation}
\mathbf{T}^{s(p)} = \mathbf{M}_{0,1}^{s(p)} \mathbf{P}_1^{s(p)} \mathbf{M}_{1,2}^{s(p)} \mathbf{P}_2^{s(p)}  \mathbf{M}_{2,3}^{s(p)} ...\mathbf{M}_{N_{lay}-1,N_{lay}}^{s(p)} \mathbf{P}_{N_{lay}}^{s(p)} \mathbf{M}_{N_{lay},{N_{lay}+1}}^{s(p)},
\end{equation}
with
\begin{align}
&\mathbf{M}_{i,i+1}^{s(p)} = \frac{1}{t_{i,i+1}^{s(p)}}
\begin{pmatrix}
1 & -r_{i,i+1}^{s(p)} \\ -r_{i,i+1}^{s(p)} & 1
\end{pmatrix}, \\
& \mathbf{P}_{i}^s = \begin{pmatrix}
e^{-i \frac{\omega}{c} \sqrt{\varepsilon_{i,y}-\xi^2}} & 0 \\ 0 & e^{i \frac{\omega}{c} \sqrt{\varepsilon_{i,y}-\xi^2} d_i}
\end{pmatrix}, \qquad  \mathbf{P}_{i}^p = \begin{pmatrix}
e^{-i \frac{\omega}{c} \sqrt{\varepsilon_{i,x}\left(1-\frac{\xi^2}{\varepsilon_{i,z}}\right)}} & 0 \\ 0 & e^{i \frac{\omega}{c} \sqrt{\varepsilon_{i,x}\left(1-\frac{\xi^2}{\varepsilon_{i,z}}\right)}}
\end{pmatrix}.
\end{align}
The matrix $\mathbf{M}_{i,i+1}^{s(p)}$ relates the amplitudes of the electric field propagating both in the $+$ and $-$ directions, in the interface between layers $i$ and $i+1$: $\begin{pmatrix}
E_{i,+}^{s(p)}\\
E_{i,-}^{s(p)}
\end{pmatrix} = \mathbf{M}_{i,i+1}^{s(p)} \begin{pmatrix}
E_{i+1,+}^{s(p)}\\
E_{i+1,-}^{s(p)}
\end{pmatrix}$. $\mathbf{P}_i^{s(p)}$ describes the propagation of light through the layer $i$. Once the total matrix $\mathbf{T}^{s(p)}= \begin{pmatrix}
T_{11}^{s(p)} & T_{12}^{s(p)} \\ T_{21}^{s(p)} & T_{22}^{s(p)}
\end{pmatrix}$ is obtained, these matrix elements are used to calculate the \textit{total} transmission and reflection coefficients of the system as $t_{total}^{s(p)} = \frac{E^{s(p)}_{N+1,+}}{E^{s(p)}_{0,+}} = \frac{1}{T_{11}^{s(p)}}$ and $r_{total}^{s(p)} = \frac{E^{s(p)}_{0,-}}{E^{s(p)}_{0,+}}=\frac{T_{21}^{s(p)}}{T_{11}^{s(p)}}$.

Last, we obtain the reflectivity spectra  $\mathcal{R}^{s(p)}$, which is defined as the ratio between the intensity of the reflected and incident light. It is obtained as $\mathcal{R}^{s(p)} = |r_{total}^{s(p)}|^2$ ($r_{total}^{s(p)}$ corresponds to the ratio of the amplitude of the electric field). We stress that, as discussed above, in every transfer-matrix simulation shown in this work except in Sec. \ref{section_focused_beam}, we consider light incident in the $z$ direction ($\xi = 0$). Under this condition, the $z$ component of the permittivity does not affect the results and, taking into account that $\varepsilon_{hBN,x}=\varepsilon_{hBN,y}$ for hBN, we recover the usual equations for isotropic materials. The anisotropy needs to be considered in Sec. \ref{section_focused_beam} for focused illumination.

\subsection{Frequencies of the modes}\label{subsec_fresnel_poles}

The coupling strength $g$ is determined from the position of the eigenfrequencies of the system. In general, however, the reflectivity dips (or transmission peaks) that are found in experimental optical spectra do not need to coincide with the eigenfrequencies, particularly for small coupling strengths. To characterize these complex eigenfrequencies, we also use the transfer-matrix formalism. From the total transfer matrix $\mathbf{T}^{s(p)}$, we obtain the analytical expression for the Fresnel coefficients $r_{total}^{s(p)}$ and $t_{total}^{s(p)}$ of the total system. Then, we solve numerically the complex frequencies for which the denominator of these coefficients vanishes, by applying all the analytical expressions for the permittivities (see Sec. \ref{sec_description}) to complex values of $\omega$.\footnote{As we use simple Lorentzian and Drude functions, the expression of the permittivities can be directly evaluated at complex frequencies.} The eigenfrequencies of the bare-cavity modes and of the polaritonic modes correspond to the complex poles of the Fresnel coefficients of the bare and hBN-filled cavities, respectively.

We compare in Fig. \ref{figure_cavity_100nm} and in Figs. 3 and 5 of the main text the real part of these eigenfrequencies (indicated by dashed lines) with the frequencies of the reflectivity dips for the three cavities explored in more detail in this work (100 nm-thick hBN layer, 10 nm-thick hBN layer and fully-filled cavity, respectively).  The agreement between these two  frequencies is very good, and thus in these systems the reflectivity dips are a good indication of the bare-cavity and polaritonic modes. However, the calculation of the eigenfrequencies is important for determining the coupling strength for very thin layers in Fig. 4 of the main text.

\subsection{Coupled harmonic oscillator model} \label{subsec_harmonic_oscillator}

In this subsection, we explain the procedure that we follow in the main text to compute the coupling strength $g$. This method is based on modelling the interaction between the material and the cavity mode by means of two coupled harmonic oscillators. This discussion is complemented in Sec. \ref{section_derivation_g} with an alternative method to calculate $g$, where we derive an analytical expression based on the microscopic interaction of the phononic material  with the electric field of the cavity mode.

In this model, one oscillator ($x_{cav}^{(j)}$) is associated to the electric field of the mode of order $j$ and frequency $\omega_{cav}^{(j)}$ of the bare cavity, where we eliminate the phonon contribution and describe the permittivity of hBN as $\varepsilon_{hBN,\infty}$. The second ($x_{TO}$) 
represents the phonon at frequency $\omega_{TO}$  (which  is modelled in Sec. \ref{section_derivation_g} explicitely as the collective excitation of the dipoles associated with each unit cell within hBN). The coupling strength between these two oscillators is $g$, and the equations of motion are \cite{novotny10, yoo20, li20, wu10}:
\begin{subequations}
\begin{align}
&\ddot{x}_{cav}^{(j)} + \kappa \dot{x}_{cav}^{(j)} + (\omega_{cav}^{(j)})^2 x_{cav}^{(j)} - 2g \dot{x}_{TO} = 0\\
&\ddot{x}_{TO} + \gamma  \dot{x}_{TO} + \omega_{TO}^2 x_{TO} + 2g \dot{x}_{cav}^{(j)} = 0.
\end{align}
\label{harmonicoscillatoreq}
\end{subequations}
$\kappa$ and $\gamma$ are the losses of the cavity and the phonon, respectively, and the dot $^\cdot$ denotes the time derivative. $\gamma = 5$ cm$^{-1}$ and $\omega_{TO} = 1364$  cm$^{-1}$ are taken from the hBN dielectric function (Sec. \ref{sec_description}) and $\omega_{cav}^{(j)}$ and $\kappa$ are extracted from the complex pole $\omega_{cav}^{(j)} - i\frac{\kappa}{2}$ of the Fresnel coefficients of the bare cavity (these eigenvalues are obtained as explained in Sec. \ref{subsec_fresnel_poles}). In our system, the losses of the cavity vary between $\kappa \approx 30 \text{ cm}^{-1}$, for the cavity fully filled with hBN, and $\kappa \approx 60 \text{ cm}^{-1}$, for cavities with very thin layers of hBN. On the other hand, the experimental measurement of this value $\kappa$ is challenging. However, by comparing the widths of the measured reflectivity dips with the simulations, we observe that the experimental widths are only slightly larger, and we estimate that the difference between the experimental and theoretical value of $\kappa$ is at most of 10$\%$ (notice that the sum of the widths of the two polaritonic dips should be close to $\kappa+\gamma$ \cite{khitrova06}).

Equation \eqref{harmonicoscillatoreq} can be alternatively written in frequency domain:
\begin{subequations}
\begin{align}
&\left[-\omega^2 - i \kappa \omega + (\omega_{cav}^{(j)})^2 \right] x_{cav}^{(j)}+2i g \omega \; x_{TO} = 0\\
&-2ig\omega \; x_{cav}^{(j)} + \left[-\omega^2 - i \gamma \omega + \omega_{TO}^2 \right] x_{TO} = 0.
\end{align}\label{oscillators_frequency_domain}
\end{subequations}
The polaritonic frequencies $\omega_{\pm}^{(j)}$ correspond to the values that cancel the determinant of this system of equations, i.e., 
\begin{equation}
\left[-(\omega_{\pm}^{(j)})^2+(\omega_{cav}^{(j)})^2-i\kappa\omega_{\pm}^{(j)}\right] \left[-(\omega_{\pm}^{(j)})^2+(\omega_{TO}^{(j)})^2-i\gamma \omega_{\pm}^{(j)} \right]-4g^2(\omega_{\pm}^{(j)})^2 = 0, \label{determinant_polariton}
\end{equation}
which can be solved numerically for any value of the coupling strength. Further, approximate analytical solutions can be obtained for two typical situations. For small values of $g$, the polaritonic frequencies are similar to the bare frequencies: $\omega_{\pm}^{(j)} \approx \omega_{cav}^{(j)}, \omega_{TO}$. Thus, we can make the secular approximation $(\omega_{cav}^{(j)})^2-(\omega_{\pm}^{(j)})^2 = \left[\omega_{cav}^{(j)}+\omega_{\pm}^{(j)} \right]\left[\omega_{cav}^{(j)}-\omega_{\pm}^{(j)} \right]\approx$ $ 2 \omega_{\pm}^{(j)} \left[\omega_{cav}^{(j)}-\omega_{\pm}^{(j)}\right]$ (and the corresponding approximation for $\omega_{TO}$) \cite{torma15}. Then, Eq. \eqref{oscillators_frequency_domain} becomes
\begin{subequations}
\begin{align}
&\left[-\omega - i \frac{\kappa}{2}  + \omega_{cav}^{(j)}\right] x_{cav}^{(j)}+i g  \; x_{TO} = 0\\
&-ig \; x_{cav}^{(j)} + \left[-\omega - i \frac{\gamma}{2}  + \omega_{TO}\right] x_{TO} = 0,
\end{align}
\end{subequations}
and  Eq. \eqref{determinant_polariton}  becomes quadratic in $\omega_{\pm}^{(j)}$, with solutions \cite{autore18}
\begin{equation}
\omega_{\pm}^{(j)} = \frac{1}{2}(\omega_{cav}^{(j)} + \omega_{TO}) \pm \frac{1}{2} \Re\left[\sqrt{\left(\omega_{cav}^{(j)}-\omega_{TO}+i \frac{\gamma-\kappa}{2}\right)^2 + 4g^2} \right].\label{solution_oscillator_strong}
\end{equation}
On the other hand, Eq. \eqref{determinant_polariton} can also be solved analytically for the limiting case where the losses $\kappa$ and $\gamma$ are negligible compared to $g$. Under this condition, the polaritonic frequencies are
\begin{equation}
\omega_{\pm}^{(j)} = \frac{1}{\sqrt{2}}\sqrt{(\omega_{cav}^{(j)})^2 + \omega_{TO}^2 + 4g^2 \pm \sqrt{\left( (\omega_{cav}^{(j)})^2 + \omega_{TO}^2 + 4g^2 \right)^2 - 4(\omega_{cav}^{(j)})^2\omega_{TO}^2}}.\label{solution_oscillator_ultrastrong}
\end{equation}

To calculate the coupling strength $g$, we first extract the polaritonic frequencies $\omega_\pm^{(j)}$ from the transfer-matrix simulations (Sec. \ref{subsec_transfer_matrix}). For this calculation,  we choose a thickness $L_{cav}$ so that the cavity mode is resonant with the TO phonon, $\omega_{cav}^{(j)}=\omega_{TO}$. Then, we obtain the value for $g$ that minimizes the difference between the two frequencies obtained from this  simulation and from the analytical solution. For the latter, we use either Eq. \eqref{solution_oscillator_strong} or Eq. \eqref{solution_oscillator_ultrastrong}, depending on the coupling regime (we have verified that the results from these approximate equations are a very good approximation of the exact Eq. \eqref{determinant_polariton} in their corresponding regimes of application). We follow this procedure to calculate all the values of $g$ given in the main text, and in particular for the analyisis of the three cavities in Fig. 2. The coupling strength that we obtain for the first cavity mode is $g = 34 \text{ cm}^{-1}$ for the cavity embedding a 10 nm-thick hBN layer, $g = 105 \text{ cm}^{-1}$ for the 100 nm-thick layer and $g = 428 \text{ cm}^{-1}$ for the cavity fully filled by hBN.

\begin{figure}
\centering
\includegraphics[scale=0.32]{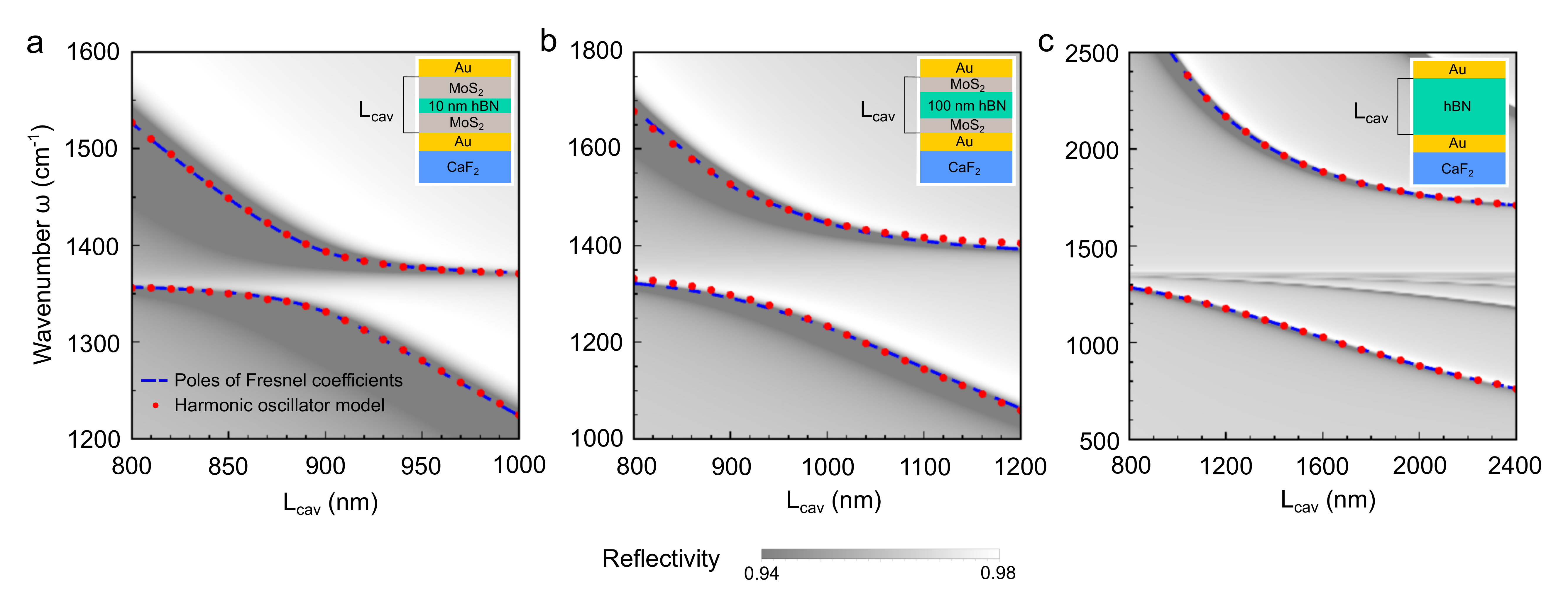}
\caption{Frequency of the polaritonic states formed by the TO phonon and the first cavity mode, for a cavity (a) filled with a 10 nm layer of hBN, (b) filled with a 100 nm layer of hBN and (c) fully filled with hBN. The colormap shows the calculated reflectivity spectra as a function of the incident wavenumber $\omega$ and the total thickness $L_{cav}$. Red dots show the polariton frequencies given by the harmonic oscillator model for (a) $g = 34 \text{ cm}^{-1}$, (b) $g = 105 \text{ cm}^{-1}$ and (c) $g = 428 \text{ cm}^{-1}$, and the blue dashed lines show the frequencies extracted from the poles of the Fresnel coefficients.} \label{figure_comparison_polaritons}
\end{figure}

\begin{figure}[H]
\centering
\includegraphics[scale=0.4]{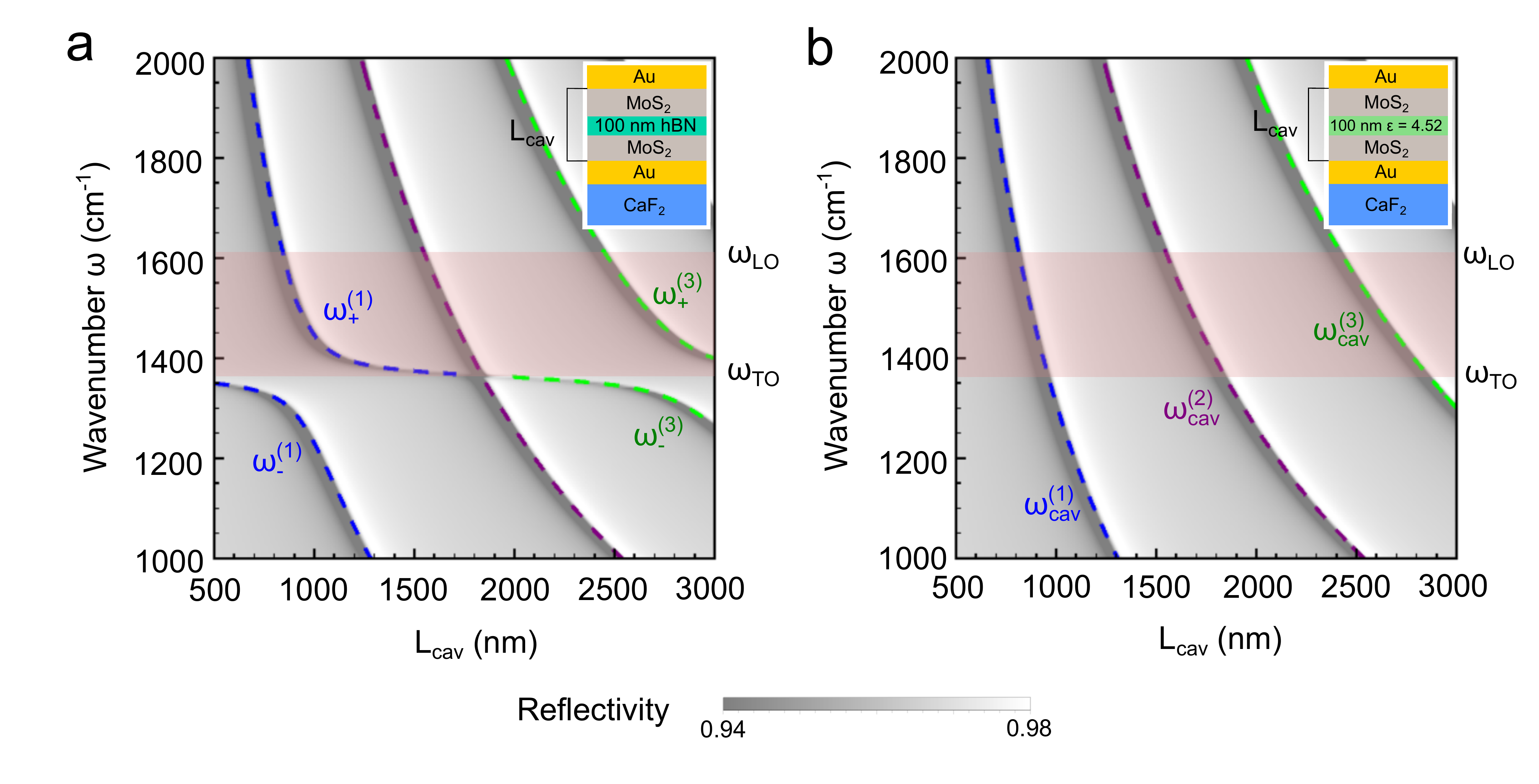}
\caption{Theoretical calculations of the reflectivity spectra of a cavity embedding a 100 nm layer of a) hBN and b) a material with the high-frequency permittivity of hBN, $\varepsilon_{hBN,\infty}$. This layer is situated between two identical MoS$_2$ slabs. The reflectivity is calculated with transfer-matrix simulations, as a function of the wavenumber of the incident light and the total thickness of the cavity $L_{cav}$. The shadowed area highlights the upper Reststrahlen band of hBN, delimited by the $\omega_{LO}$ and $\omega_{TO}$  wavenumbers. The dashed lines represent the  wavenumbers of the polariton modes $\omega_{+}^{(j)}$ and $\omega_{-}^{(j)}$ in (a) and the bare cavity modes $\omega_{cav}^{(j)}$ in (b), as obtained from the poles of the Fresnel coefficients.}\label{figure_cavity_100nm}
\end{figure}

Further, although we calculated $g$ for the particular case where the cavity is tuned to the TO phonon, we show in Fig. \ref{figure_comparison_polaritons} that these values of $g$  allow to fully describe the evolution of the polaritonic frequencies as the cavity is detuned. Specifically, Fig. \ref{figure_comparison_polaritons} shows the polaritonic frequencies obtained from the harmonic oscillator model with (a) $g = 34 \text{ cm}^{-1}$, (b) $g = 105 \text{ cm}^{-1}$ and (c) $g = 428 \text{ cm}^{-1}$ (red dots) compared with those given by the poles of the Fresnel coefficients for  (a) a 10 nm-thick hBN layer, (b) a 100 nm-thick hBN layer, and  (c) a fully-filled cavity (dashed blue lines), as a function of the total cavity thickness $L_{cav}$. These values are superimposed to the colormap of the calculated reflectivity of the cavity. The frequencies obtained from these two methods are almost identical for all $L_{cav}$ and for the three  cavities considered.

Interestingly, Eq. \eqref{solution_oscillator_ultrastrong} coincides with the eigenvalues of the Hopfield Hamiltonian $\hat{H}_{hop}$ \cite{hopfield58}. In this alternative approach, light and matter excitations are also represented by harmonic oscillators which interact with coupling strength $g$, but the behaviour of the system is described by the Hamiltonian
\begin{equation}
\hat{H}_{hop} = \hbar \omega_{cav}^{(j)} \hat{a}^\dagger \hat{a} + \hbar \omega_{TO} \hat{b}^\dagger \hat{b} + \hbar g \sqrt{\frac{\omega_{TO}}{\omega_{cav}^{(j)}}} (\hat{a} + \hat{a}^\dagger)(\hat{b} + \hat{b}^\dagger) + \hbar \frac{g^2}{\omega_{cav}^{(j)}}(\hat{a} + \hat{a}^\dagger)^2,
\end{equation}
where $\hat{a}$ and $\hat{a}^\dagger$ are the annihilation and creation operator of the $j^{th}$ cavity mode, respectively, while $\hat{b}$ and $\hat{b}^\dagger$ are the operators for the TO phonon. Apart from the interaction term proportional to $g$, this Hamiltonian also includes the diamagnetic term, which is proportional to $g^2$ \cite{kockum19, deliberato14}. The fact that this Hamiltonian leads to the same eigenfrequencies as given by Eq. \eqref{solution_oscillator_ultrastrong} indicates that the classical harmonic oscillator model is able to describe the polaritonic frequencies also in the ultrastrong coupling regime \cite{yoo20}.

Last, we note that the harmonic oscillator model presented here treats  each cavity mode of order $j$ independently, i.e. that  different cavity modes do not interact with each other and thus the coupling between hBN and each cavity mode is described by a separate Eq. \eqref{harmonicoscillatoreq}. For the cavities that we analyze in this work, which are designed so that the TO phonon interacts mostly with the first cavity mode, the results in Fig. \ref{figure_comparison_polaritons} show that this approach is appropriate. However, we have found that for thicker cavities where higher-order modes are tuned to the TO phonon, the harmonic oscillator does not always describe well the position of the polaritonic frequencies as obtained from the transfer-matrix simulations. We believe that solving this discrepancy requires to include in the harmonic oscillator model the possibility of coupling different modes with each other, for instance following an approach based on quasinormal modes \cite{sauvan13, franke19} (a similar reason may be behind the small discrepancy between the two values of the coupling strength that are obtained in Fig. \ref{figure_comparison_g} for the 3$^{\text{rd}}$ cavity mode, as discussed in Sec. \ref{section_derivation_g}). However, the results of the cavities analyzed in this work would not be affected by future analysis of this type.

\section{Reflectivity spectra of the cavity embedding a 100 nm-thick hBN layer}

In the main text, we show the reflectivity spectra of a cavity embedding a 10 nm-thick hBN layer (Fig. 3) and a cavity fully filled by hBN (Fig. 5), as a function of the total cavity thickness $L_{cav}$. To complement these results, we show in Fig. \ref{figure_cavity_100nm}a the reflectivity spectra of the cavity embedding a 100 nm-thick hBN layer. For reference,  in Fig. \ref{figure_cavity_100nm}b we show a calculation of a bare cavity where the 100 nm layer is formed by a material with constant permittivity $\varepsilon_{hBN,\infty}$. The dashed lines in Fig. \ref{figure_cavity_100nm}a indicate the polaritonic frequencies $\omega_-^{(j)}$ and $\omega_+^{(j)}$ that result from the coupling between the TO phonon and the bare-cavity mode of order $j$. The dashed lines in Fig.  \ref{figure_cavity_100nm}b correspond to the bare cavity mode frequencies $\omega_{cav}^{(j)}$. As expected, we observe that the anticrossing between the polaritonic frequencies  for the 100 nm-thick hBN layer is considerably smaller than the fully-filled cavity, but much larger than the cavity with a 10 nm layer.

\begin{figure}[H]
\centering
\includegraphics[scale=0.25]{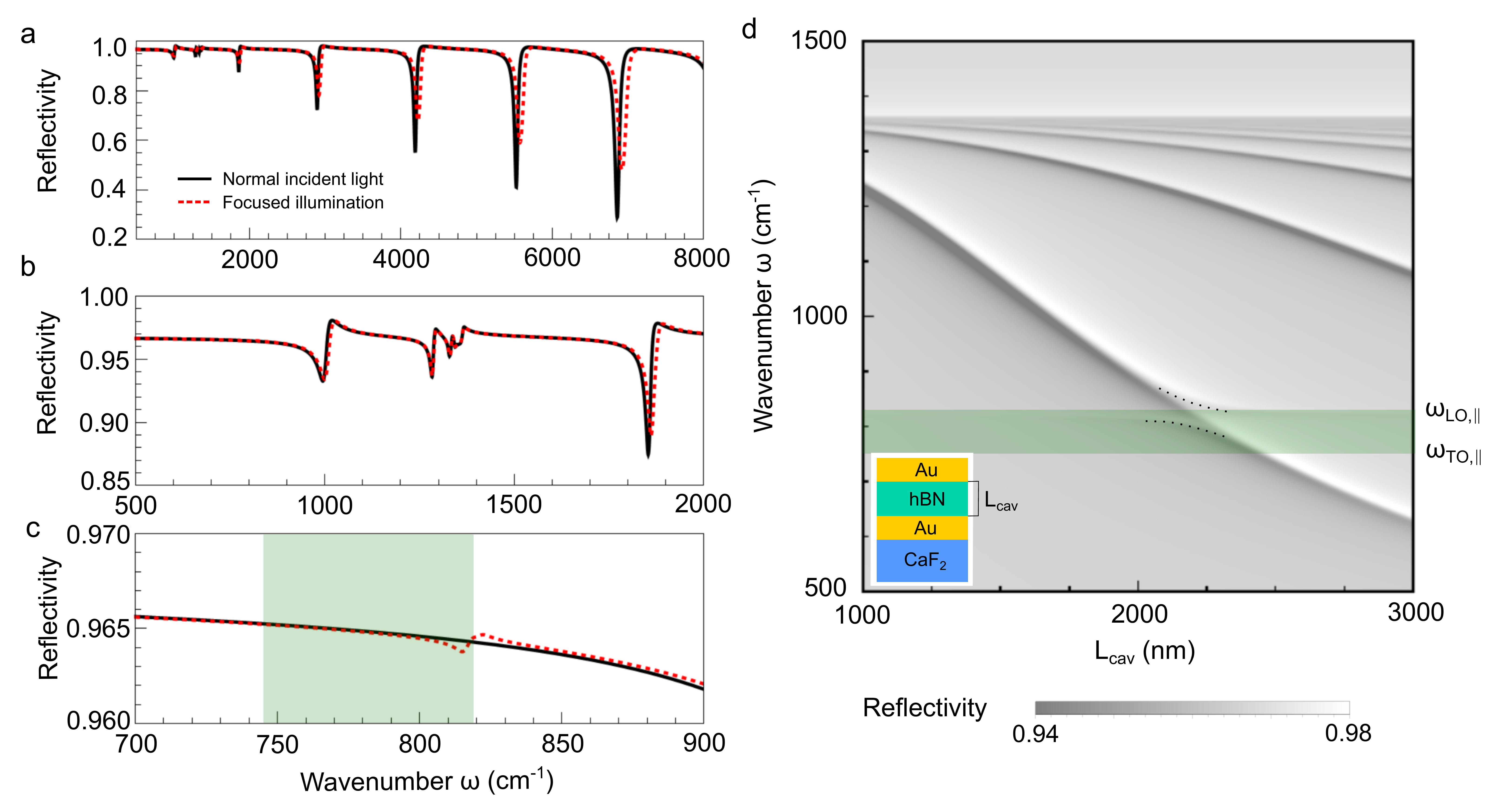}
\caption{Reflectivity spectra calculated for cavities fully filled with hBN under focused illumination. a-c) Comparison between the spectra obtained for normal incident light (black solid lines) and focused illumination (red dashed lines). (b) and (c) correspond to zooms of different regions of the spectra in (a). d) Reflectivity spectra under focused illumination as a function of the total cavity thickness $L_{cav}$ and  wavenumber $\omega$ of the incident light. Dots are a guide to the region near the anticrossing that occurs due to the coupling of the cavity mode of order 1 and the out-of-plane phonons. The green shadowed area in (c) and (d) represents the lower Reststrahlen band of hBN, limited by frequencies $\omega_{TO,\parallel}$ and $\omega_{LO,\parallel}$.} \label{figure_focused_beam}
\end{figure} 

\section{Reflectivity of the system under focused illumination} \label{section_focused_beam}

In the transfer-matrix calculations of the main text, we assume that the propagation direction of the incident light is normal to the surface of the mirrors and the substrate ($z$ direction). However, the experiments have been performed using a microscope with focused illumination. In this section,  we discuss the effect of this focusing on the reflectivity spectra of the system. We show that this effect is small and that the main difference with respect to the spectra under normal incident light is the possibility to observe the coupling with the out-of-plane phonons of hBN, instead of just with the in-plane phonons (see Sec. \ref{sec_description} for a discussion of these two types of phonons).

The total reflectivity $\mathcal{R}_{foc}$ for focused illumination is obtained by decomposing the incident light as an integral over planewaves incident at different angles. We then obtain (for more details, see Chapter 3.9 in Ref. \cite{novotny-hecht}):
\begin{equation}
\mathcal{R}_{foc} = \frac{I_{ref}}{I_{in}} = \frac{\int_{\sin \theta_{min}}^{\sin{\theta_{max}}} \frac{|r_{total}^s(\xi)|^2+|r_{total}^p(\xi)|^2}{2}  \; \xi \; d\xi}{\int_{\sin \theta_{min}}^{\sin{\theta_{max}}} \; \xi \; d\xi},\label{R_focal}
\end{equation}
where $I_{in}$ and $I_{ref}$ correspond to the intensity of the incident and reflected focused beam, respectively, $\xi = \frac{\sqrt{k_x^2+k_y^2}}{k_0} = \frac{k_\parallel}{k_0}$ is the normalized parallel wavevector and $r_{total}^{s(p)}$ is the $\xi$-dependent  Fresnel reflection coefficient of the full system for a $s(p)$-polarized planewave, which is discussed thoroughly in Sec. \ref{subsec_transfer_matrix}. The upper limit of the integral is given by the numerical aperture of the microscope  used to focus light, $\text{NA} = \sin{\theta_{max}}$. In our case, we have $\text{NA}=0.4$, and thus, $\theta_{max} \approx 23.5^{\circ}$. Furthermore, we set $\theta_{min} = 10^{\circ}$ for the lower bound of the integral, because the microscope used in the experiments obstructs the propagation of the central part of the light incoming to the focusing lens, which eliminates the contribution from the small-angle components.

We have verified that, for the cavity embedding a 10 nm layer of hBN, the results of the transfer-matrix simulations using focused illumination are nearly identical compared to the spectra for normal incidence (not shown). Crucially, light travels at considerably smaller angles inside this cavity than in free space, due to the high permittivity of MoS$_2$, $\varepsilon_{MoS2}$, which fills most of the cavity for such a thin hBN layer. In particular, the angles inside MoS$_2$ range from $2.6^{\circ}$ to $5.9^{\circ}$, and thus the difference with the case of normal incidence is very small.

We next consider a fully-filled hBN cavity of thickness $L_{cav} = 1665$ nm (whose first cavity mode is resonant with the TO phonon frequency) and show in Fig. \ref{figure_focused_beam}(a-c) the reflectivity spectra under focused (red dashed line)  and  normal-incidence illumination (black solid line), both calculated by transfer-matrix simulations. The effect of using focused light becomes larger for the cavity fully filled by hBN, as the high-frequency permittivity of hBN $\varepsilon_{hBN,\infty}$ is smaller than $\varepsilon_{MoS2}$, and thus the angle of light propagation inside the cavity can be larger.
A small but appreciable difference between normal incident light and focused illumination is observed for high wavenumbers, as shown in panel a. The reflectivity dips obtained under focused illumination are displaced towards larger wavenumbers compared to the normal-incidence spectra, because the frequencies of the Fabry-Pérot cavity modes depend on the  angle of propagation. Furthermore, the widths of the dips are larger for focused illumination because they are the result of the sum of different contributions, each corresponding to a different angle and thus resonant at a slightly different wavenumber (Eq. \eqref{R_focal}). However, these differences remain small, and the agreement between the two spectra improves further for frequencies close to $\omega_{TO}$ (see zoom in panel b), which is the main region of interest in this work. Thus, the use of normal incidence in all our calculations of the other sections is justified.

Possibly the most interesting feature of the focused illumination is observed in Fig. \ref{figure_focused_beam}c, which shows the reflectivity spectra at low wavenumbers (near the frequencies of the out-of-plane phonons). An additional reflectivity dip for focused illumination (red dashed line) appears compared to the normal-incidence spectra (black solid line), which is due to the anisotropy of hBN. For normal incidence, the illumination only couples with the in-plane transverse optical phonon, which is polarized in the parallel direction to all planar interfaces. On the other hand, for focused illumination it becomes possible to excite the phonons that are polarized in the out-of-plane direction and that are found at significantly lower frequencies:  $\omega_{TO,\parallel} = 746 \text{ cm}^{-1}$ and $\omega_{LO,\parallel} = 819 \text{ cm}^{-1}$ (the lower Reststrahlen band limited by these frequencies is highlighted with the green area in Fig. \ref{figure_focused_beam}). These phonons strongly affect the $z$ component of the permittivity tensor, and thus they can couple with the electric field components in the $z$ direction of the focused light, which explains the extra dip.  

For a more detailed analysis of this coupling,  we show in Fig. \ref{figure_focused_beam}d the calculated reflectivity spectra under focused illumination as a function of the total cavity thickness $L_{cav}$. When the frequency of the bare-cavity mode gets close to the lower Reststrahlen band, an anticrossing characterized by a relatively small  Rabi splitting is observed, as indicated by the dots as a guide. Hence, the use of a focused beam makes possible to observe the coupling of the cavity modes with in-plane and out-of-plane phonons.

\begin{figure}
\centering
\includegraphics[scale=0.35]{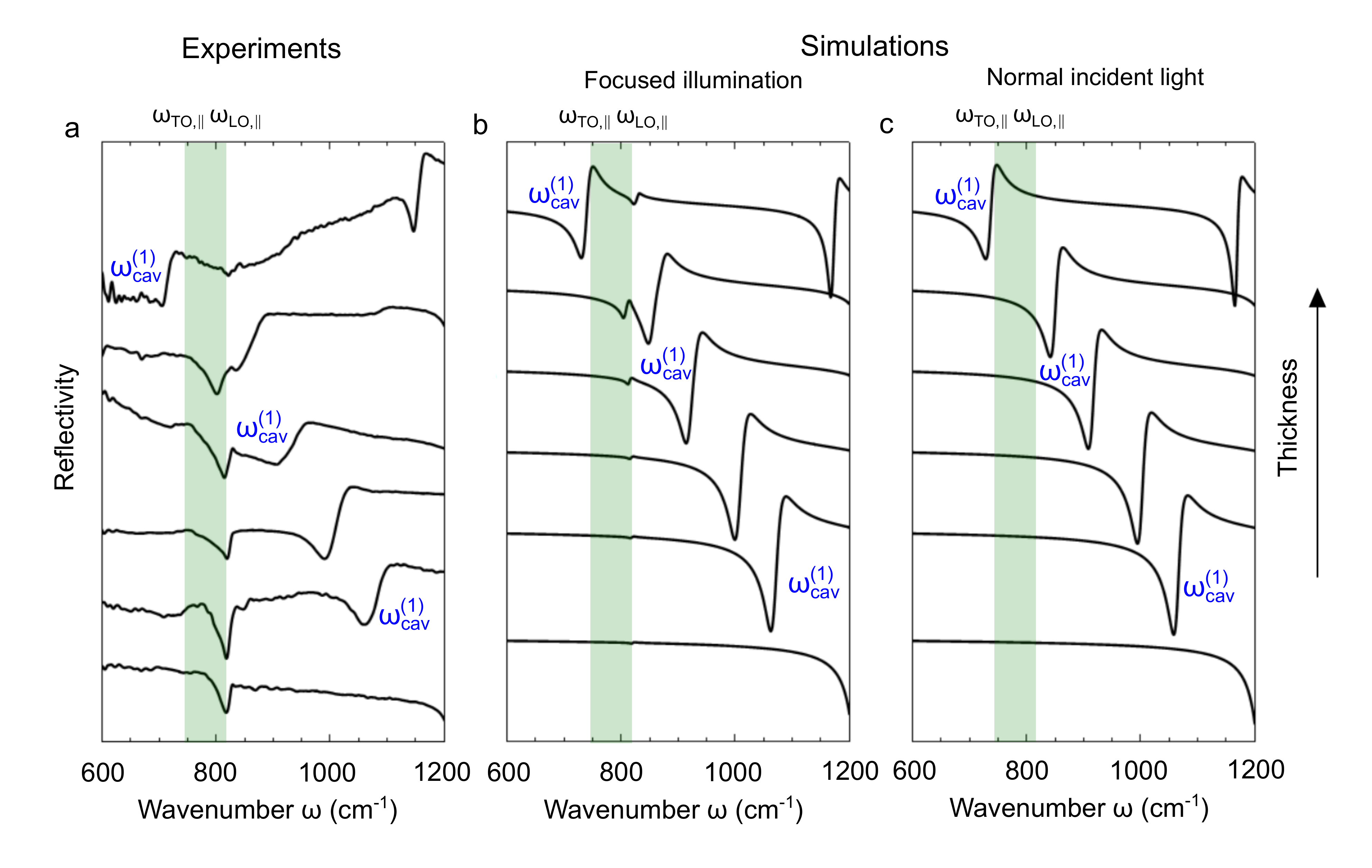}
\caption{Reflectivity spectra of the fully-filled hBN cavities at low wavenumbers, obtained from a) the experiments; b) transfer-matrix simulations under focused illumination; and c) transfer-matrix simulations with a planewave incoming in the normal direction. The lower Reststrahlen band is highlighted by the green area. The (a) nominal and (b,c) simulated thicknesses $L_{cav}$ of the cavities  (from bottom to top) are: 1080 nm, 1500 nm, 1665 nm, 1900 nm, 2100 nm and 2500 nm. Panel a is a zoom of Fig. 5c of the main text in the region of low wavenumbers.} \label{figure_focused_beam_experiment}
\end{figure} 

The coupling with the out-of-plane phonons has also been observed in the experimental spectra of the fully-filled hBN cavities.  Figure \ref{figure_focused_beam_experiment}a  shows the measured reflectivity spectra for all six cavities. This figure corresponds to a zoom of Fig. 5c in the main text  on the region near the lower Reststrahlen band (green area) associated with the out-of-plane phonons. We observe a dip labelled by $\omega_{cav}^{(1)}$ that changes strongly with the frequency and it is  mostly associated to the bare-cavity mode. The other dip is close to $\omega_{LO,\parallel}$, and its frequency is also indicated by the yellow dots in Fig. 5b of the main text. In order to confirm the nature of the dips at frequency $\approx \omega_{LO,\parallel}$, we also show in Fig. \ref{figure_focused_beam_experiment} the results from the transfer-matrix  simulations under focused illumination (panel b) and normal incident light (panel c). For panel c, we only see the dip associated to the bare-cavity mode at frequency $\omega_{cav}^{(1)}$. On the other hand, once we consider focused illumination, we can observe the second feature near $\omega_{LO,\parallel}$ that was identified  in the experiments. We thus confirm that the extra peak only appears in the calculations when the polarization has a nonzero $z$ component and thus the illumination couples also with the out-of-plane phonons. However, we notice that the size of the experimental dips is significantly larger than the simulated ones, and we attribute this discrepancy to experimental imperfections such as rugosities (which can scatter light at high angles) or non-perfect planarity of the fabricated cavities. Despite this  difference, the good agreement on the spectral positions of the dips indicates that they are indeed a result of the coupling between the first cavity mode and the out-of-plane phonons.

\section{Analytical expression for the coupling strength}\label{section_derivation_g}

\subsection{Coupling strength of a TO phonon with a Fabry-Pérot cavity mode}

In Sec. \ref{subsec_harmonic_oscillator}, we describe the method that we follow in the main text to calculate the coupling strength $g$ for hBN cavities, by extracting the polaritonic frequencies from the poles of the Fresnel coefficients and fitting these frequencies with a model of coupled harmonic oscillators. In this section, we present an alternative approach based on the microscopic interaction between the phononic material and the modes of a Fabry-Pérot cavity, which allows for deriving an approximate analytical expression for $g$, and can give additional physical insight about the coupling.

In this derivation, we consider a simplified picture where we model each unit cell of hBN as a harmonic oscillator characterized by a transition dipole moment $\mathbf{d_i}$. Thus, from this perspective, the material is a collection of $N_{cell}$ dipoles interacting with the cavity mode. Due to the homogeneity of the material, the transition dipole moments of all unit cells are the same: $\mathbf{d_i} = \mathbf{d}$. All dipoles are oriented parallel to the electric field,  and in order to deduce the module $d_i=|\mathbf{d_i}|$, we first consider the electric susceptibility of the material, which for a polar material such as hBN is
\begin{equation}
\chi(\omega) = \varepsilon_\infty \frac{\omega_{LO}^2-\omega_{TO}^2}{\omega_{TO}^2-\omega^2-i\omega \gamma}.\label{susceptibility_hBN}
\end{equation}
$\omega_{TO}$ and $\omega_{LO}$ are the frequencies of the transverse optical and the longitudinal optical phonon, respectively, $\gamma$ is the damping frequency and $\varepsilon_{\infty}$ is the high-frequency permittivity. From this expression, we can directly obtain the polarizability $\alpha_i$ of each dipole induced in a unit cell of volume $V_{cell}$. We
consider that the polarization density $\mathbf{P}$ is related to the electric field $\mathbf{E}$  as $\mathbf{P} = \frac{\alpha_i}{V_{cell}} \mathbf{E} = \varepsilon_0  \chi \mathbf{E}$, and we therefore obtain
\begin{equation}
\alpha_i(\omega) = \varepsilon_0  V_{cell}\chi(\omega),\label{polarizability_hBN}
\end{equation}
where $\varepsilon_0$ is the vacuum permittivity. We can now relate between the transition dipole moment $d_i$ and the classical polarizability by \cite{novotny-hecht}
\begin{equation}
\alpha_i(\omega) = \frac{2\omega_{TO}}{\hbar}\frac{d_i^2}{\omega_{TO}^2-\omega^2-i\omega\gamma}, \label{polarizability_TLS}
\end{equation}
and, from Eqs. \eqref{susceptibility_hBN}--\eqref{polarizability_TLS}:
\begin{equation}
d_i = \sqrt{\frac{\hbar}{2\omega_{TO}}V_{cell}\varepsilon_0 \varepsilon_{\infty}(\omega_{LO}^2-\omega_{TO}^2)}.
\end{equation}

This expression allows to characterize the coupling strength $g_i$ between the dipole associated to a particular unit cell  and the cavity mode with quantized electric field $\hat{\mathbf{E}}(\mathbf{r})$. This electric field corresponds to a bare cavity filled by a material of frequency-independent, real permittivity $\varepsilon_\infty$ (i.e., without considering the resonant polarizability of the unit cells). For the quantization of the field, we further assume that the mirrors of the cavity are perfect, so that the electric and magnetic energy of the modes are equal and the fields do not penetrate into the mirrors. The electric field of the Fabry-Pérot mode of order $j$ resonant at frequency $\omega_{cav}^{(j)}$  is quantized as \cite{todorov12}
\begin{equation}
\mathbf{\hat{E}} = \mathbf{E}(\hat{a}+\hat{a}^\dagger) = \sqrt{\frac{\hbar \omega_{cav}^{(j)}}{2\varepsilon_0 V_{eff}}} \rho(z)(\hat{a} + \hat{a}^\dagger)\mathbf{\hat{u}_x}.
\end{equation}
$\hat{a}$ and $\hat{a}^\dagger$ are the annihilation and creation operators of the cavity mode, respectively. The fields only vary along the $z$ direction normal to the flat interfaces of the system, as given by the field profile $\rho(z)$ (normalized so that $\rho(z)=1$ in the position where the field amplitude is maximum). The electric field is assumed to be polarized along the $x$ axis, with unit vector $\mathbf{\hat{u}_x}$. $V_{eff} = S \int \varepsilon(z) |\rho(z)|^2 dz$ is the effective volume of the field, where $S$ is the effective surface of the cavity and $\varepsilon(z)$ indicates the spatial distribution of the permittivities of the system. The value of the electric field $\mathbf{E}(\mathbf{r})$ of a Fabry-Pérot mode is related to the coupling strength $g_i$ between that mode and the dipole excited in the unit cell at position $\mathbf{r_i}$, according to
\begin{equation}
\hbar g_i = - \mathbf{d_i} \cdot \mathbf{\mathbf{E}}(\mathbf{r_i}).\label{definition_g}
\end{equation}
Taking into account that the dipole moments and the electric field are parallel, we write this coupling strength as
\begin{equation}
g_i = -\sqrt{\frac{\omega_{cav}^{(j)}}{\omega_{TO}}\frac{\omega_{LO}^2-\omega_{TO}^2}{4}\frac{L_{cell}\varepsilon_{\infty} |\rho(z_i)|^2}{ \int \varepsilon(z) |\rho(z)|^2 dz}}, \label{g_unitcell}
\end{equation}
where $L_{cell} = \frac{V_{cell}}{S}$ is the length of the unit cell along the $z$ axis.

We next consider the Hamiltonian describing the interaction of the cavity mode with the ensemble of dipoles associated with all the unit cells in the phononic material
\begin{equation}
\hat{H} = \hbar \omega_{cav}^{(j)} \hat{a}^\dagger \hat{a} + \hbar \omega_{TO} \sum_{i=1}^{N_{cell}} \hat{b}^\dagger_i \hat{b}_i + \hbar(\hat{a}+\hat{a}^\dagger)\sum_{i=1}^{N_{cell}} (g_i \hat{b}_i + g_i^* \hat{b}^\dagger_i).
\end{equation}
where $\hat{b}_i$ and $\hat{b}_i^\dagger$ are the annihilation and creation operator of the harmonic oscillator that represents each vibrational dipole, respectively. The next step is to write the Hamiltonian of the system in terms of the annihilation $\hat{b}_c$ and creation $\hat{b}_c^\dagger$ operators of the collective excitation that represents the TO phonon,
\begin{equation}
\hat{H} = \hbar \omega_{cav}^{(j)} \hat{a}^\dagger \hat{a} + \hbar \omega_{TO} \hat{b}_c^\dagger \hat{b}_c + \hbar g(\hat{a}+\hat{a}^\dagger)(\hat{b}_c+\hat{b}_c^\dagger),
\end{equation}
which can be done using the Dicke transformation
\begin{equation}
\sum_{i=1}^{N_{cell}} \hat{b}^\dagger_i \hat{b}_i = \hat{b}_c^\dagger \hat{b}_c, \qquad \sum_{i=1}^{N_{cell}} g_i \hat{b}_i = g \hat{b}_c, \qquad \sum_{i=1}^{N_{cell}} g_i^* \hat{b}_i^\dagger = g^* \hat{b}_c^\dagger,
\end{equation}
where $g$ takes the value \cite{zhang20}
\begin{equation}
g = \sqrt{\sum_{i=1}^{N_{cell}} |g_i|^2}\label{sum_collective_g}
\end{equation}
so that the commutation relation $[b_c,b^\dagger_c] = 1$ is fulfilled. Thus, this expression gives the coupling strength between the cavity mode and the collective excitation (the TO phonon) in terms of the coupling strengths $g_i$ associated to the individual dipoles of each unit cell. $g$ is evaluated by replacing Eq. \eqref{g_unitcell} into Eq. \eqref{sum_collective_g}, and transforming the sum into an integral (since the dipoles are distributed in a continuous way). The result is
\begin{align}
|g|^2 =& \sum_{i=1}^{N_{cell}} \frac{\omega_{cav}^{(j)}}{\omega_{TO}}\frac{\omega_{LO}^2-\omega_{TO}^2}{4}\frac{L_{cell}\varepsilon_{\infty} |\rho(z_i)|^2}{\int \varepsilon(z') |\rho(z')|^2 dz'} = \int \frac{\omega_{cav}^{(j)}}{\omega_{TO}}\frac{\omega_{LO}^2-\omega_{TO}^2}{4}\frac{L_{cell}\varepsilon_{\infty} |\rho(z)|^2}{\int \varepsilon(z') |\rho(z')|^2 dz'} \frac{d{N_{cell}}}{dz}dz \nonumber \\
=& \frac{\omega_{cav}^{(j)}}{\omega_{TO}} \frac{\omega_{LO}^2-\omega_{TO}^2}{4} \frac{\int \varepsilon_{\infty} |\rho(z)|^2 dz}{\int \varepsilon(z') |\rho(z')|^2 dz'},\label{relation_collective_g}
\end{align}
where in the last step we have taken into account that the density of dipoles per unit length is $\frac{d{N_{cell}}}{dz} = \frac{1}{L_{cell}}$. Therefore, in this subsection we have obtained an expression that allows to calculate the coupling strength between the TO phonon of a polar material and a Fabry-Pérot mode of a cavity with an arbitrary  spatial distribution of the permittivity $\varepsilon(z)$ (and thus arbitrary field distribution $\rho(z)$).

\subsection{Application to the hBN microcavities}

 In this subsection, we focus on the particular system that we analyze in the main text, a cavity of thickness $L_{cav}$ with MoS$_2$ as a spacer and containing a layer of hBN (details can be found in Sec. \ref{sec_description}).
First, we need to calculate the field profile $\rho(z)$ of the modes of the bare cavity, in order to evaluate the coupling strength $g$ through Eq. \eqref{relation_collective_g}. Labelling the positions of the MoS$_2$-hBN interfaces as $L_1$ and $L_2$  (indicated in Fig. \ref{figure_cavity_diagram}), the spatial distribution of the permittivity $\varepsilon(z)$ of the bare cavity is described by the function
\begin{equation}
\varepsilon(z) = 
\begin{cases}
\varepsilon_{MoS_2}, \qquad &0 < z < L_1 \\
\varepsilon_{hBN,\infty}, \qquad &L_1 < z < L_2 \\
\varepsilon_{MoS_2}, \qquad &L_2 < z < L_{cav} \\
\end{cases} .
\end{equation}
In each interval with constant permittivity $\varepsilon_i$, the field profile $\rho(z)$ satisfies the Helmholtz equation
\begin{equation}
\frac{d^2 \rho(z)}{dz^2}+ \frac{\varepsilon_i (\omega_{cav}^{(j)})^2}{c^2}\rho(z) = 0,
\end{equation}
where $c$ is the speed of light in vacuum. Further, since the cavity has been assumed to be made of perfect mirrors, the electric field vanishes at both ends: $\rho(0) = \rho(L_{cav}) = 0$. In order to verify the boundary conditions, $\rho(z)$ needs to be continuous and differentiable in all interfaces, which leads to the solution
\begin{equation}
\rho(z) = 
\begin{cases}
A\frac{\sin(\frac{\omega_{cav}^{(j)}}{c}\sqrt{\varepsilon_{hBN,\infty}}L_1+\phi^{(j)})}{\sin(\frac{\omega_{cav}^{(j)}}{c}\sqrt{\varepsilon_{MoS_2}}L_1)} \sin(\frac{\omega_{cav}^{(j)}}{c}\sqrt{\varepsilon_{MoS_2}}z), \qquad &0 < z < L_1\\
A\sin(\frac{\omega_{cav}^{(j)}}{c}\sqrt{\varepsilon_{hBN,\infty}}z+\phi^{(j)}), \qquad &L_1 < z < L_2\\
A\frac{\sin(\frac{\omega_{cav}^{(j)}}{c}\sqrt{\varepsilon_{hBN,\infty}}L_1+\phi^{(j)})}{\sin(\frac{\omega_{cav}^{(j)}}{c}\sqrt{\varepsilon_{MoS_2}}L_1)} \sin(\frac{\omega_{cav}^{(j)}}{c}\sqrt{\varepsilon_{MoS_2}}(z-L_{cav})), \qquad &L_2 < z < L_{cav}
\end{cases}.
\end{equation}
$A$ is the normalization constant chosen so that the maximum of the field profile is $\rho(z)=1$, while $\omega_{cav}^{(j)}$ and $\phi^{(j)}$ are the $j^{th}$ solution of the system of equations
\begin{subequations}
\begin{align}
&\sqrt{\varepsilon_{MoS_2}}\cot(\frac{\omega_{cav}^{(j)}}{c}\sqrt{\varepsilon_{MoS_2}}L_1) = \sqrt{\varepsilon_{hBN,\infty}}\cot(\frac{\omega_{cav}^{(j)}}{c}\sqrt{\varepsilon_{hBN,\infty}}L_1+\phi^{(j)})\\
&\sqrt{\varepsilon_{MoS_2}}\cot(\frac{\omega_{cav}^{(j)}}{c}\sqrt{\varepsilon_{MoS_2}}(L_2-L_{cav})) = \sqrt{\varepsilon_{hBN,\infty}}\cot(\frac{\omega_{cav}^{(j)}}{c}\sqrt{\varepsilon_{hBN,\infty}}L_2+\phi^{(j)}).
\end{align}
\end{subequations}

We can  obtain $g$ using Eq. \eqref{relation_collective_g}, with the integral extending over the phononic material (between $L_1$ and $L_2$). In this calculation, we always assume that the cavity mode of order $j$ is resonant with the TO phonon, $\omega_{TO} = \omega_{cav}^{(j)}$, so that the cavity thickness $L_{cav}$ is different for each order and the permittivity of MoS$_2$ is evaluated at that frequency ($\varepsilon_{MoS_2} \equiv  \varepsilon_{MoS_2}(\omega_{TO})$ in this subsection).  We obtain 
\begin{equation}
|g|^2 = \frac{\omega_{LO}^2-\omega_{TO}^2}{4}\frac{\int_{L_1}^{L_2} \varepsilon_{hBN,\infty} |f(z)|^2 dz}{\int_{0}^{L_{cav}} \varepsilon(z') |f(z')|^2 dz'}.\label{finalformula_g}
\end{equation}
Adjusting the cavity parameters (specially the thickness of the hBN layer) allows for varying $g$ from 0 to a maximum value $g_{max}$ corresponding to the fully-filled cavity that can be evaluated to be $g_{max} = \sqrt{\frac{\omega_{LO}^2-\omega_{TO}^2}{4}}$. Interestingly, we observe that $g_{max}$ only depends on the frequencies of the transverse optical phonon and the longitudinal optical phonon. For hBN, $g_{max} = 428 \text{ cm}^{-1} $.

\begin{figure}
\centering
\includegraphics[scale=0.4]{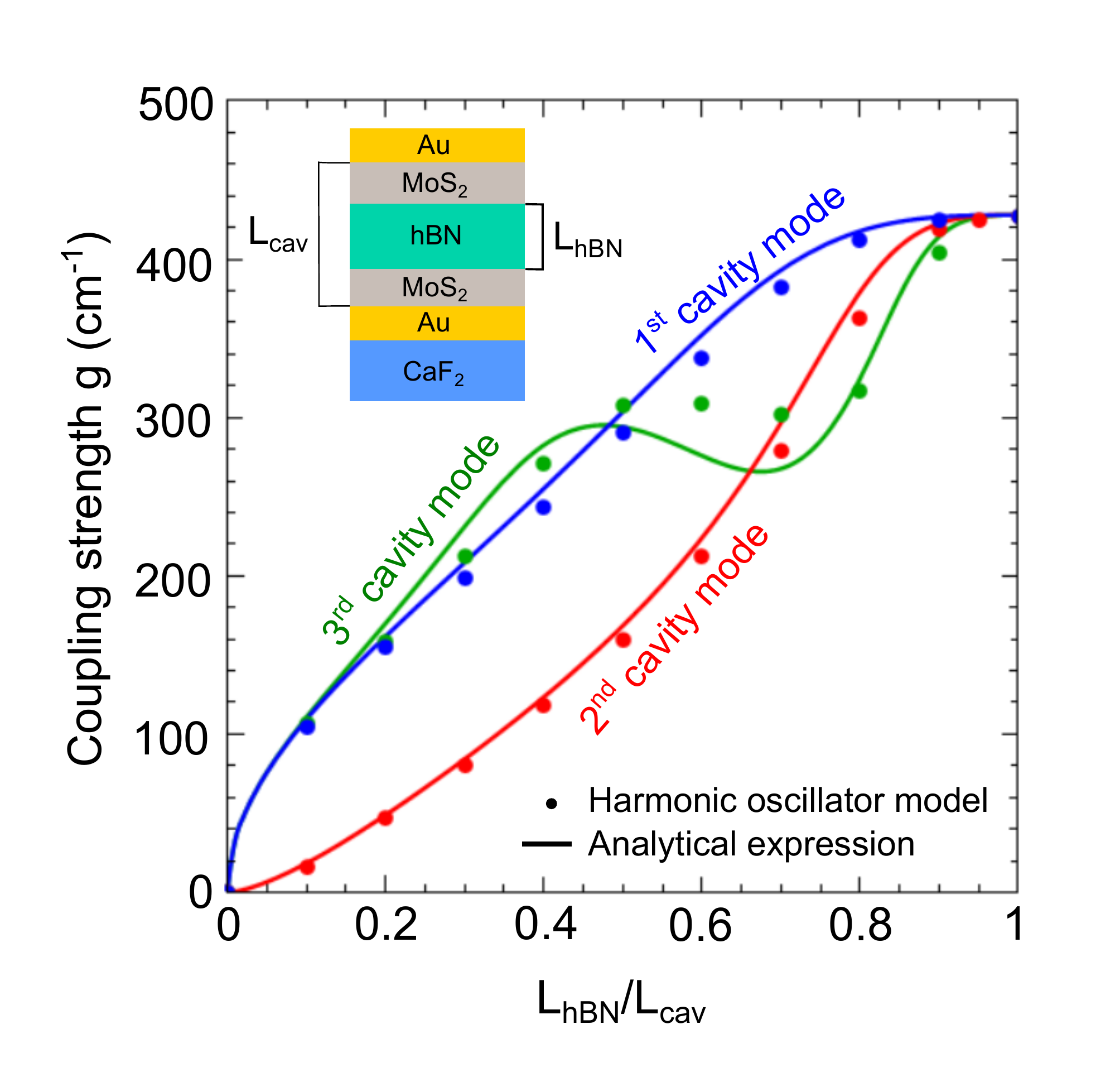}
\caption{Evolution of the coupling strength $g$ as a function of the ratio $\frac{L_{hBN}}{L_{cav}}$ (shown in the inset) for the first (blue), the second (red) and the third (green) cavity modes. Solid lines represent the solution of Eq. \eqref{finalformula_g}, and the dots correspond to the values obtained from the coupled harmonic oscillator model (blue and red dots correspond to the values shown in Fig. 4 of the main text). The cavity thickness $L_{cav}$ is changed so that the cavity is always resonant with the  TO phonon frequency.} \label{figure_comparison_g}
\end{figure} 

Further, we can also obtain a simple analytical expression of $g$ for thin layers of hBN placed in the middle of the cavity and interacting with the first cavity mode. In order to obtain the field profile $\rho(z)$ in this regime, we consider that the thin layer of hBN disturbs very weakly the electric field of the cavity fully filled with MoS$_2$ and with the same total thickness $L_{cav}$. Under this assumption, we obtain that $\rho(z) \approx \sin\left( \frac{\omega_{cav}^{(1)}}{c} \sqrt{\varepsilon_{MoS_2}}z \right)$, with $\omega_{cav}^{(1)} =  \frac{\pi c}{\sqrt{\varepsilon_{MoS_2}}L_{cav}} $. Moreover, since the thin layer is located in the position of the maximum amplitude of the electric field, this amplitude varies slowly inside the hBN layer, and we can assume that the field is constant between $L_1$ and $L_2$ $\left(\rho(z) \approx 1 \right)$. By evaluating the integrals in Eq. \eqref{finalformula_g} for $L_1 = \frac{L_{cav}}{2}-\frac{L_{hBN}}{2}$ and $L_2 = \frac{L_{cav}}{2}+\frac{L_{hBN}}{2}$, we obtain
\begin{equation}
g \approx \sqrt{ \frac{\omega_{LO}^2-\omega_{TO}^2}{2}\frac{\varepsilon_{hBN,\infty}}{\varepsilon_{MoS_2}} \frac{L_{hBN}}{L_{cav}}} \approx 332 \text{ cm}^{-1} \sqrt{\frac{L_{hBN}}{L_{cav}}} .\label{g_thinlayers}
\end{equation}
This result is consistent with the Dicke model \cite{dicke54, garraway11} that has been often applied to quantify the coupling strength between a cavity mode and an ensemble of $N$ identical molecules \cite{kaluzny83,raizen89,thompson92}. Since in our case the thickness  of the hBN layer $L_{hBN}$ is proportional to the amount of dipoles interacting  with the cavity mode, we obtain the same dependency of $g$ on the amount of matter excitations $g\propto \sqrt{N}$ as predicted by Dicke.

We assess in Fig. \ref{figure_comparison_g} the validity of our analytical model by comparing the result of Eq. \eqref{finalformula_g} with the coupling strengths calculated by applying a coupled harmonic oscillator model to the polaritonic frequencies calculated with the transfer-matrix method. For the calculation of the polaritonic frequencies, we  always choose the thickness $L_{cav}$ so that the frequency of the cavity mode is resonant with the TO phonon frequency. We show the evolution of the coupling strength $g$ as a function of the filling factor $\frac{L_{hBN}}{L_{cav}}$ for the first three cavity modes. Generally, $g$ increases with the filling factor, particularly strongly for thin layers and odd modes, until for all modes it saturates at the same value $g_{max}$ for fully-filled cavities. We also find behavior that can be non-intuitive such as a decrease of the coupling strength with increased $\frac{L_{hBN}}{L_{cav}}$ (for the third mode in the interval $ 0.5\lesssim \frac{L_{hBN}}{L_{cav}} \lesssim 0.7$), or an increase faster than expected from Eq. \eqref{g_thinlayers}. The reason is that changing  $L_{hBN}$ also modifies $L_{cav}$ (to keep the cavity resonant with the TO phonon) and the field distribution inside the cavity, which leads to a complex behavior. Further, for the coupling strength of the first two modes, we observe that the agreement between the analytical expression (solid lines) and the harmonic oscillator model (dots) is very good, with only a slight overestimation by the analytical equation for intermediate values of the filling factor. The agreement becomes worse for the third cavity mode and intermediate filling factors, but it remains generally reasonable. Thus, we confirm that the simple analytical model presented here is able to explain the evolution of the coupling strength with the filling factor, particularly for the first two cavity modes considered in the main text.

\end{document}